\documentclass[twocolumn,trackchanges]{aastex63}
\usepackage{fp}
\usepackage{bm}
\usepackage{xcolor}
\DeclareSymbolFont{matha}{OML}{txmi}{m}{it}
\DeclareMathSymbol{\varv}{\mathord}{matha}{118}

\usepackage{multirow}

\shorttitle{{\it HST} Study of the Offset and Host Light Distributions of SLSNe}
\shortauthors{Hsu et al.}

\newcommand{\STSci}{\affiliation{Space Telescope Science Institute, 3700 San Martin Dr, Baltimore, MD 21218, USA}}

\newcommand{\CfA}{\affiliation{Center for Astrophysics \textbar{} Harvard \& Smithsonian, 60 Garden Street, Cambridge, MA 02138-1516, USA}}
\newcommand{\UA}{\affiliation{Steward Observatory, University of Arizona, 933 North Cherry Avenue, Tucson, AZ 85721-0065, USA}}

\newcommand{\CIERA}{\affiliation{Center for Interdisciplinary Exploration and Research in Astrophysics and Department of Physics and Astronomy, \\Northwestern University, 1800 Sherman Avenue 8th Floor, Evanston, IL 60201, USA}}

\newcommand{\IAIFI}{\affiliation{The NSF AI Institute for Artificial Intelligence and Fundamental Interactions}}

\defcitealias{Planck_2020}{Planck Collaboration 2018}
\defcitealias{Gaia_2016}{Gaia Collaboration 2016}
\defcitealias{Gaia_2022}{Gaia Collaboration 2022}
\defcitealias{Lunnan_2015}{L15}
\defcitealias{Blanchard_2016}{B16}
\begin{document}

\title{An Extensive {\it Hubble Space Telescope} Study of the Offset and Host Light Distributions of Type I Superluminous Supernovae}

\correspondingauthor{Brian Hsu}
\email{bhsu@arizona.edu}

\author[0000-0002-9454-1742]{Brian~Hsu}
\CfA
\UA

\author[0000-0003-0526-2248]{Peter~K.~Blanchard}
\CIERA

\author[0000-0002-9392-9681]{Edo~Berger}
\CfA
\IAIFI

\author[0000-0001-6395-6702]{Sebastian~Gomez}
\STSci

\begin{abstract}
We present an extensive {\it Hubble Space Telescope} ({\it HST}) rest-frame ultraviolet (UV) imaging study of the locations of Type I superluminous supernovae (SLSNe) within their host galaxies. The sample includes 65 SLSNe with detected host galaxies in the redshift range $z\approx 0.05-2$. Using precise astrometric matching with SN images, we determine the distributions of physical and host-normalized offsets relative to the host centers, as well as the fractional flux distribution relative to the underlying UV light distribution. We find that the host-normalized offsets of SLSNe roughly track an exponential disk profile, but exhibit an overabundance of sources with large offsets of $1.5-4$ times their host half-light radius. The SLSNe normalized offsets are systematically larger than those of long gamma-ray bursts (LGRBs), and even Type Ib/c and II SNe.  Furthermore, we find that about 40\% of all SLSNe occur in the dimmest regions of their host galaxies (fractional flux of 0), in stark contrast to LGRBs and Type Ib/c and II SNe.  We do not detect any significant trends in the locations of SLSNe as a function of redshift, or as a function of explosion and magnetar engine parameters inferred from modeling of their optical lights curves. The significant difference in SLSN locations compared to LGRBs (and normal core-collapse SNe) suggests that at least some of their progenitors follow a different evolutionary path. We speculate that SLSNe arise from massive runaway stars from disrupted binary systems, with velocities of $\sim 10^2$ km s$^{-1}$.
\end{abstract}

\section{Introduction}
\label{introduction}

Hydrogen-poor (Type I) superluminous supernovae (hereafter, SLSNe; \citealt{Chomiuk_2011,Quimby_2011,Gal-Yam_2012}) are a rare subclass of core-collapse supernovae (CCSNe) that are $\sim 10-100$ times more luminous than normal SNe and have longer durations of several months to years (e.g. \citealt{Nicholl_2015b,Inserra_2017,Lunnan_2018,DeCia_2018}). Originally defined to have a peak absolute magnitude of $M<-21$ (\citealt{Gal-Yam_2012}), they are now classified based on their spectra, which are dominated by a blue continuum devoid of hydrogen features, and typically exhibit distinctive early-time, ``W''-shaped \ion{O}{2} absorption lines at $\sim 3600-4600$ \AA\ (\citealt{Chomiuk_2011,Quimby_2011,Lunnan_2013,Mazzali_2016,Quimby_2018}). The volumetric rate of SLSNe is only $\sim 0.1\%$ of the overall CCSN rate \citep{Quimby_2018,Frohmaier_2021}, but in magnitude-limited optical surveys they account for $\sim 2\%$ of all transients (e.g., \citealt{Villar_2019,Perley_2020,Gomez_2021}) thanks to their large luminosity. 

Shortly after their discovery, it became clear that SLSNe are not powered by the radioactive decay of $^{56}$Ni (\citealt{Arnett_1982}) as in Type Ib/c SNe (SNe Ib/c). This is due to the unusually large mass of radioactive nickel required to match the observed luminosity, often exceeding the total ejecta mass, and in conflict with the lack of suppression of their UV emission due to line blanketing from iron-peak elements (e.g., \citealt{Dessart_2012,Inserra_2013,Nicholl_2013,Liu_2017,Nicholl_2017a,Nicholl_2017b,Yan_2017a,Gal-Yam_2019}). Instead, alternative mechanisms have been proposed as the main power source of SLSNe, including a central engine with a rapidly spinning ($\sim$ few ms) and highly magnetized ($\sim 10^{14}$ G) neutron star (magnetar model; \citealt{Kasen_Bildsten_2010,Woosley_2010,Dessart_2012,Metzger_2015,Nicholl_2017b}), or shock interaction with a hydrogen-poor circumstellar medium (CSM model; \citealt{Chevalier_Irwin_2011,Chatzopoulos_2012}).

The magnetar model has been highly successful in accounting for the broad range of peak luminosities and timescales (e.g., \citealt{Nicholl_2017b,Blanchard_2020,Hsu_2021}), for the early UV/optical spectra (e.g., \citealt{Nicholl_2017c}), for the nebular phase spectra (e.g., \citealt{Nicholl_2016b,Nicholl_2019,Jerkstrand_2017}), and for the power law decline rates observed in SN\,2015bn and SN\,2016inl at $\gtrsim 10^3$ days \citep{Nicholl_2018,Blanchard_2021}. On the other hand, there are a handful of events supporting CSM interaction as the dominant power source for SLSNe based on the presence of interaction lines (e.g., \citealt{Yan_2015,Yan_2017b}) and equally well-explained light curves (e.g., \citealt{Chatzopoulos_2012,Chen_2022b}). However, the spectroscopic properties exhibited by the majority of SLSNe (e.g., \citealt{Pastorello_2010,Quimby_2011,Quimby_2018,Inserra_2013,Liu_2017}) and limits placed by X-ray/radio observations (\citealt{Margutti_2018,Eftekhari_2021}) are hard to reconcile with the CSM model alone. 

In addition to the power source of SLSNe, it is critical to explore and constrain possible progenitor systems. The lack of hydrogen spectral features in SLSNe point to a connection with the stripped-envelope massive star progenitors of SNe Ib/c (e.g., \citealt{Pastorello_2010}). However, it has been shown with a uniformly modeled light curve sample that SLSN progenitors are systematically more massive ($\approx 3.6-40\ \rm M_{\odot}$; \citealt{Blanchard_2020}) than SNe Ib/c progenitors ($\approx3.7-5.4\ \rm M_{\odot}$; \citealt{Barbarino_2020}). Comparing the SLSN progenitor mass distribution at the time of explosion to stellar evolutionary models suggests that low-metallicity binary systems may be plausible progenitors (e.g., \citealt{Liu_2015,Moriya_2015,Blanchard_2020,Stevance_2021}).

Studies of SLSN host galaxies have also been leveraged to shed light on their progenitors.  These studies revealed a preference for low-metallicity dwarf galaxies with higher specific star formations rates (sSFRs) and lower luminosity than CCSNe (e.g., \citealt{Chen_2013,Lunnan_2013,Lunnan_2014,Lunnan_2015,Perley_2016,Angus_2016,Schulze_2018}). An investigation of the sub-galactic environments of 16 SLSNe using high-resolution {\it HST} data suggested that SLSN locations track bright UV regions of their host galaxies (\citealt{Lunnan_2015}; hereafter, \citetalias{Lunnan_2015}), consistent with massive progenitors. The relatively small sample size, however, did not allow for a statistically meaningful comparison with other classes of transients such as long gamma-ray bursts (LGRBs) and CCSNe.

With about 150 spectroscopically confirmed SLSNe to date (\citealt{Gomez_2020,Chen_2022a}), it is now possible to significantly expand on the early analysis of SLSN locations (\citetalias{Lunnan_2015}).  Here, we take advantage of a large set of archival {\it HST} SLSN host galaxy rest-frame UV observations from a wide range of programs, to produce the first large (65 SLSNe) and statistically meaningful sample of SLSN sub-galactic locations.  We follow the same methodology that has been employed to explore the progenitors of other populations of transients, including SNe Ia (\citealt{Wang_2013,Anderson_2015}), SNe Ib/c and II  (\citealt{Kelly_2008,Prieto_2008,Svensson_2010,Kelly_Kirshner_2012}), LGRBs (\citealt{Bloom_2002,Fruchter_2006,Svensson_2010,Blanchard_2016}; hereafter \citetalias{Blanchard_2016}), and short GRBs \citep{Fong_2013}.

The paper is structured as follows. We present the sample, {\it HST} imaging, data processing techniques, and astrometric matching to determine the SLSN locations in \S\ref{sec:sample}. In \S\ref{sec:methods} we describe our measurement methodology for determining host associations, offsets, and fractional fluxes. In \S\ref{sec:results} we present the resulting offset and fractional flux distributions and compare these to other transients. In \S\ref{sec:discussion} we explore trends and correlations with redshift, as well as with inferred SLSN explosion and magnetar engine properties, and discuss implications for SLSN progenitors. We conclude with a summary of our findings in \S\ref{sec:conclusion}.

Throughout the paper, we assume a flat $\Lambda$CDM cosmology with $\Omega_\mathrm{m}=0.310$ and $H_0=67.7\ \text{km}\ \text{s}^{-1}\ \text{Mpc}$, based on the Planck 2018 results \citepalias{Planck_2020}. All observations are reported in AB magnitudes (\citealt{Oke_Gunn_1983}) and corrected for Galactic extinction using \citet{Schlafly_Finkbeiner_2011}, following the \citet{Gordon_2023} extinction law with $R_V=3.1$.

\section{Data and Astrometry} 
\label{sec:sample}

\subsection{{\it HST} Data}
\label{sec:HST_data}

\begin{figure}[t!]
    \centering
    \includegraphics[width=\columnwidth]{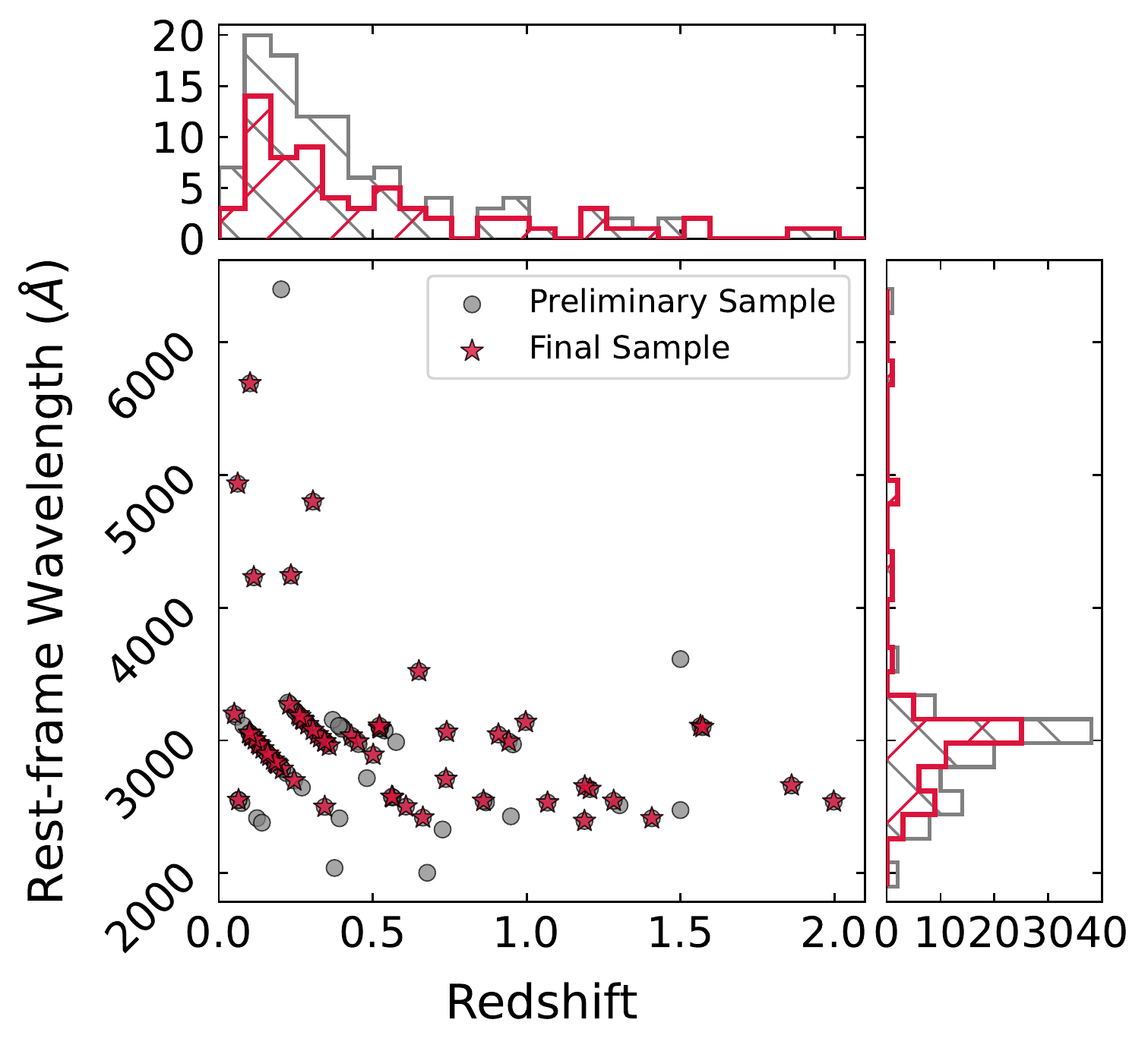}
    \caption{Rest-frame effective wavelengths probed by the {\it HST} datat as a function of the SLSN redshifts. Grey points denote the full sample, while red points indicate the final sample used in the analysis (see \S\ref{sec:sample}). We note that only 5/65 SLSN hosts in the final sample are not observed in the rest-frame UV. The top and right panels show the projected distributions of redshift and rest-frame wavelength, respectively.}
    \label{fig:redshift}
\end{figure}

We compiled an initial sample of 109 Type I SLSNe with {\it HST} host galaxy observations from archival and ongoing programs, imaged with either the Advanced Camera for Surveys (ACS) or the Wide Field Camera 3 (WFC3).  The names, redshifts, and details of the {\it HST} observations are listed in Table~\ref{tab:objects}. The initial sample spans a wide redshift range of $z=(0.05-1.998)$, where the {\it HST} filters used in the observations predominantly probe a rest-frame wavelength range in the UV of $\approx 2300-3300$\ \AA; see Figure~\ref{fig:redshift}. This allows us to probe the locations of the SLSNe relative to the underlying star formation activity in their hosts.

\startlongtable
\begin{deluxetable*}{llcccccccccc}
\tablewidth{\textwidth} 
\tablecaption{Prelimary Sample of SLSNe with Available {\it HST} Host Galaxy Images}
\tablehead{
\colhead{} & \colhead{} & \multicolumn{5}{c}{{\it HST} Image}& \multicolumn{2}{c}{SLSN Image}\\[-10pt]
\colhead{} & \colhead{} & \multicolumn{5}{c}{--------------------------------------------------------------------------------------------}& \multicolumn{2}{c}{---------------------------------}\\[-10pt]
\colhead{} & \colhead{} & \colhead{} & \colhead{} & \colhead{Exp. Time} & \colhead{Obs. Date} & \colhead{Program} & \colhead{Telescope/} & \colhead{}\\[-12pt]
\colhead{SLSN} & \colhead{Redshift} & \colhead{Instrument} & \colhead{Filter} & \colhead{} & \colhead{} & \colhead{} & \colhead{} & \colhead{Filter}\\[-13pt]
\colhead{} & \colhead{} & \colhead{} & \colhead{} & \colhead{(s)} & \colhead{(UT)} & \colhead{ID} & \colhead{Instrument} & \colhead{}}

\startdata
DES13S2cmm & 0.663 & WFC3/UVIS2 & F475W & 1500 & 2023 Feb 08 & 17181 & Blanco/DECam & $i$ \\
DES14C1fi & 1.302 & WFC3/UVIS & F606W & 2500 & 2018 Jul 05 & 15303 & \nodata & \nodata \\
DES14C1rhg & 0.481 & WFC3/UVIS & F390W & 5610 & 2018 Aug 23 & 15303 & Blanco/DECam & $g$ \\
DES14S2qri & 1.50 & WFC3/UVIS & F625W & 5576 & 2018 Dec 16 & 15303 & Blanco/DECam & $i$ \\
DES14X2byo & 0.868 & WFC3/UVIS & F475W & 5582 & 2018 Feb 01 & 15303 & Blanco/DECam & $r$ \\
DES14X3taz & 0.608 & WFC3/UVIS & F390W & 2456 & 2018 Jul 15 & 15303 & Blanco/DECam & $z$ \\
DES15C3hav & 0.392 & WFC3/UVIS & F336W & 2480 & 2018 Feb 25 & 15303 & \nodata & \nodata \\
DES15E2mlf & 1.861 & WFC3/UVIS & F775W & 2564 & 2017 Nov 22 & 15303 & Blanco/DECam & $i$ \\
DES15S1nog & 0.565 & WFC3/UVIS & F390W & 2456 & 2018 Feb 02 & 15303 & Blanco/DECam & $i$ \\
DES15S2nr & 0.22 & WFC3/UVIS & F336W & 2456 & 2017 Dec 28 & 15303 & Blanco/DECam & $i$ \\
DES15X1noe & 1.188 & WFC3/UVIS & F555W & 2476 & 2018 Jul 08 & 15303 & Blanco/DECam & $i$ \\
DES15X3hm & 0.86 & WFC3/UVIS & F475W & 5582 & 2018 Feb 06 & 15303 & Blanco/DECam & $i$ \\ 
DES16C2aix & 1.068 & WFC3/UVIS & F555W & 2500 & 2019 Jan 30 & 15303 & Blanco/DECam & $i$ \\
DES16C2nm & 1.998 & WFC3/UVIS & F775W & 2488 & 2018 Sep 01 & 15303 & Blanco/DECam & $z$ \\
DES16C3cv & 0.727 & WFC3/UVIS & F390W & 5610 & 2017 Nov 30 & 15303 & Blanco/DECam & $z$ \\
DES16C3dmp & 0.562 & WFC3/UVIS & F390W & 2480 & 2018 Jul 03 & 15303 & Blanco/DECam & $i$\\
DES16C3ggu & 0.949 & WFC3/UVIS & F475W & 2500 & 2018 Aug 30 & 15303 & \nodata & \nodata \\
iPTF13ajg & 0.7403 & ACS/WFC1 & F555W & 5290 & 2017 Sep 11 & 15140 & P48/MOSAIC & $r$ \\
iPTF13bdl & 0.403 & ACS/WFC1 & F435W & 2180 & 2018 Dec 29 & 15140 & \nodata & \nodata \\
iPTF13bjz & 0.271 & WFC3/UVIS2 & F390W & 2472 & 2018 Apr 29 & 15140 & P48/MOSAIC & $r$ \\
iPTF13cjq & 0.396 & ACS/WFC1 & F435W & 2188 & 2017 Oct 20 & 15140 & \nodata & \nodata \\
iPTF13dcc & 0.431 & ACS/WFC1 & F435W & 2172 & 2018 Jan 26 & 15140 & P48/MOSAIC & $r$ \\
iPTF13ehe & 0.3434 & WFC3/UVIS2 & F390W & 2804 & 2017 Dec 18 & 15140 & P48/MOSAIC & $r$ \\
iPTF14dck & 0.576 & ACS/WFC1 & F475W & 2180 & 2017 Sep 11 & 15140 & \nodata & \nodata \\
iPTF14dek & 0.332 & WFC3/UVIS2 & F390W & 2476 & 2018 Aug 13 & 15140 & \nodata & \nodata\\
iPTF14tb & 0.942 & WFC3/UVIS2 & F555W & 1500 & 2023 Jan 29 & 17181 & P48/MOSAIC & $r$ \\
iPTF15cyk & 0.539 & WFC3/UVIS2 & F475W & 1500 & 2023 Feb 12 & 17181 & \nodata & \nodata \\
iPTF16bad & 0.2467 & WFC3/UVIS2 & F390W & 1500  & 2023 Jan 10 & 17181 & \nodata & \nodata\\
PS1-10awh & 0.908 & ACS/WFC1 & F606W & 680 & 2013 Sep 04 & 13022 & PS1/GPC1 & $i$ \\
PS1-10bzj & 0.65 & ACS/WFC & F606W & 2160 & 2002 Nov 11 & 9500 & PS1/GPC1 & $i$\\
PS1-10ky & 0.956 & ACS/WFC1 & F606W & 680 & 2013 Dec 13 & 13022 & PS1/GPC1 & $i$\\
PS1-10pm & 1.206 & ACS/WFC1 & F606W & 1960 & 2012 Dec 10 & 13022 & PS1/GPC1 & $z$ \\
PS1-11afv & 1.407 & ACS/WFC1 & F606W & 1960 & 2013 Apr 09 & 13022 & PS1/GPC1 & $i$\\
PS1-11aib & 0.997 & ACS/WFC1 & F625W & 1000 & 2013 Sep 12 & 12529 & {\it HST}/ACS/WFC1 & F625W \\
PS1-11ap & 0.524 & ACS/WFC1 & F475W & 2464 & 2013 Oct 09 & 13326 & PS1/GPC1 & $i$\\
PS1-11bam & 1.565 & ACS/WFC1 & F814W & 2304 & 2013 Oct 11 & 13326 & PS1/GPC1 & $i$\\
PS1-11bdn & 0.738 & ACS/WFC1 & F475W & 2200 & 2013 Nov 13 & 13326 & PS1/GPC1 & $g$\\
PS1-11tt & 1.283 & ACS/WFC1 & F606W & 1960 & 2012 Dec 02 & 13022 & PS1/GPC1 & $i$\\
PS1-12bmy & 1.572 & ACS/WFC1 & F814W & 2224 & 2013 Sep 17 & 13326 & PS1/GPC1 & $i$\\
PS1-12bqf & 0.522 & ACS/WFC1 & F475W & 2200 & 2013 Nov 18 & 13326 & PS1/GPC1 & $i$\\
PS1-14bj & 0.5215 & WFC3/UVIS2 & F475W & 1500 & 2023 Jan 29 & 17181 & PS1/GPC1 & $i$\\
PTF09as & 0.186 & WFC3/UVIS2 & F336W & 2476 & 2018 Jul 18 & 15140 & P48/MOSAIC & $r$ \\
PTF09atu & 0.501 & ACS/WFC1 & F435W & 5204 & 2017 Sep 11 & 15140 & P48/MOSAIC & $r$ \\
PTF09cnd & 0.259 & WFC3/UVIS & F390W & 2224 & 2012 Nov 11 & 13025 & P48/MOSAIC & $r$ \\
PTF10aagc & 0.207 & WFC3/UVIS2 & F336W & 2468 & 2017 Oct 15 & 15140 & P48/MOSAIC & $r$ \\
PTF10bfz & 0.169 & WFC3/UVIS2 & F336W & 2468 & 2018 Apr 25 & 15140 & P48/MOSAIC & $r$ \\
PTF10bjp & 0.359 & WFC3/UVIS2 & F390W & 2804 & 2017 Dec 29 & 15140 & P48/MOSAIC & $r$ \\
PTF10hgi & 0.098 & WFC3/UVIS2 & F336W & 5570 & 2018 Sep 22 & 15140 & P48/MOSAIC & $r$ \\
PTF10nmn & 0.124 & WFC3/UVIS2 & F336W & 2456 & 2018 Feb 23& 15140 & P48/MOSAIC & $r$ \\
PTF10uhf & 0.2882 & WFC3/UVIS2 & F390W & 2600 & 2018 Aug 19 & 15140 & P48/MOSAIC & $r$ \\
PTF10vqv & 0.452 & ACS/WFC1 & F435W & 2172 & 2018 Jan 11 & 15140 & P48/MOSAIC & $r$ \\
PTF11hrq & 0.057 & WFC3/UVIS2 & F336W & 920 & 2014 Nov 14 & 13858 & P48/MOSAIC & $r$ \\
PTF12dam & 0.108 & WFC3/UVIS2 & F336W & 984 & 2014 Oct 16 & 13858 & P48/MOSAIC & $r$ \\
PTF12gty & 0.177 & WFC3/UVIS2 & F336W & 2468 & 2018 Jul 03 & 15140 & P48/MOSAIC & $r$ \\
PTF12hni & 0.1056 & WFC3/UVIS2 & F336W & 2456 & 2017 Oct 15 & 15140 & P48/MOSAIC & $r$\\
PTF12mxx & 0.327 & WFC3/UVIS2 & F390W & 5622 & 2017 Oct 24 & 15140 & P48/MOSAIC & $r$ \\
SCP06F6 & 1.189 & ACS/WFC1 & F606W & 8054 & 2013 May 23 & 13025 & {\it HST}/ACS/WFC1 & F775W \\
SN 2005ap & 0.283 & WFC3/UVIS & F390W & 1804 & 2012 Nov 25 & 13025 & \nodata & \nodata \\
SN 2006oz & 0.396 & WFC3/UVIS2 & F300X & 1200 & 2017 Sep 02 & 14762 & \nodata & \nodata \\
SN 2007bi & 0.128 & WFC3/UVIS & F336W & 1808 & 2012 Nov 27 & 13025 & LT/RATCam & $r$ \\
SN 2009jh & 0.349 & WFC3/UVIS & F390W & 2044 & 2012 Dec 06 & 13480 & P48/MOSAIC & $r$ \\
SN 2010gx & 0.23 & WFC3/UVIS & F390W & 1808 & 2012 Nov 22 & 13025 & Gemini-S/GMOS & $r$\\
SN 2010hy & 0.19 & WFC3/UVIS2 & F336W & 5592 & 2017 Oct 24 & 15140 & \nodata & \nodata \\
SN 2011ke & 0.385 & WFC3/UVIS & F336W & 2044 & 2013 May 16 & 13025 & P48/MOSAIC & $r$ \\
SN 2011kf & 0.245 & WFC3/UVIS & F336W & 2036 & 2013 Jun 29 & 13025 & LT/RATCam & $i$ \\
SN 2011kg & 0.192 & WFC3/UVIS & F336W & 1804 & 2012 Nov 12 & 13025 & P48/MOSAIC & $r$ \\
SN 2011kl & 0.677 & WFC3/UVIS & F336W & 1050 & 2011 Dec 20 & 12786 & \nodata & \nodata \\
SN 2012il & 0.175 & WFC3/UVIS & F336W & 2036 & 2013 Jan 02 & 13025 & \nodata & \nodata \\
SN 2015bn & 0.114 & ACS/WFC1 & F475W & 2344 & 2017 Jun 01 & 14743 & {\it HST}/ACS/WFC1 & F475W \\
SN 2016ard & 0.2025 & ACS/WFC1 & F775W & 2120 & 2019 Mar 19 & 15496 & \nodata & \nodata \\
SN 2016eay & 0.1013 & ACS/WFC1 & F625W & 2180 & 2019 Mar 26  & 15162 & {\it HST}/ACS/WFC1 & F625W \\
SN 2016inl & 0.3057 & ACS/WFC1 & F625W & 2140 & 2019 Jul 06 & 15162 & {\it HST}/ACS/WFC1 & F625W \\
SN 2016wi & 0.224 & WFC3/UVIS2 & F390W & 1500 & 2022 Dec 18 & 17181 & \nodata & \nodata \\
SN 2017ens & 0.1086 & WFC3/UVIS2 & F336W & 1500 & 2023 Jan 20 & 17181 & \nodata & \nodata \\
SN 2018avk & 0.132 & WFC3/UVIS2 & F336W & 1500 & 2023 Mar 05 & 17181 & P48/MOSAIC & $g$\\
SN 2018bgv & 0.0795 & WFC3/UVIS2 & F336W & 1500 & 2022 Nov 16 & 17181 & FLWO/KeplerCam & $i$\\
SN 2018bym & 0.274 & WFC3/UVIS2 & F390W & 1500 & 2022 Dec 05 & 17181 & P48/MOSAIC & $r$ \\
SN 2018don & 0.073 & WFC3/UVIS2 & F275W & 1200 & 2022 Jan 09 & 16657 & P48/MOSAIC & $r$ \\
SN 2018fcg & 0.344 & WFC3/UVIS2 & F336W & 1500 & 2022 Nov 21 & 17181& P48/MOSAIC & $r$ \\
SN 2018fd & 0.263 & WFC3/UVIS2 & F390W & 1500 & 2023 Jan 19 & 17181 & FLWO/KeplerCam & $i$\\
SN 2018ffj & 0.234 & WFC3/UVIS & F555W & 710 & 2021 Sep 26 & 16239 & P48/MOSAIC & $r$ \\
SN 2018hpq & 0.124 & WFC3/UVIS2 & F275W & 1200 & 2022 Feb 17 & 16657 & P48/MOSAIC & $r$ \\
SN 2018hti & 0.0612 & WFC3/UVIS & F555W & 710 & 2021 Oct 01 & 16239 & P48/MOSAIC & $r$ \\
SN 2018kyt & 0.108 & WFC3/UVIS2 & F336W & 1500 & 2022 Dec 05 & 17181 & P48/MOSAIC & $g$ \\
SN 2018lzv & 0.434 & WFC3/UVIS2 & F438W & 1500 & 2023 Mar 05 & 17181 & \nodata & \nodata \\
SN 2018lzw & 0.3198 & WFC3/UVIS2 & F390W & 1500 & 2022 Dec 15 & 17181 & P48/MOSAIC & $r$ \\
SN 2019cdt & 0.153 & WFC3/UVIS2 & F336W & 1500 & 2023 Feb 21 & 17181 & P48/MOSAIC & $r$ \\
SN 2019eot & 0.3057 & WFC3/UVIS2 & F390W & 1500 & 2022 Dec 02 & 17181 & P48/MOSAIC & $r$ \\
SN 2019kcy & 0.40 & WFC3/UVIS2 & F438W & 1500 & 2023 Jan 20 & 17181 & P48/MOSAIC & $g$ \\
SN 2019lsq & 0.14 & WFC3/UVIS2 & F336W & 1500 & 2023 Jan 06 & 17181 & P48/MOSAIC & $r$ \\
SN 2019neq & 0.1059 & WFC3/UVIS2 & F336W & 1500 & 2023 Feb 19 & 17181 & P48/MOSAIC & $r$ \\
SN 2019nhs & 0.19 & WFC3/UVIS2 & F336W & 1500 & 2022 Dec 31 & 17181 & P48/MOSAIC & $g$\\
SN 2019sgh & 0.344 & WFC3/UVIS2 & F390W & 1500 & 2022 Nov 19 & 17181 & FLWO/KeplerCam & $i$\\
SN 2019stc & 0.117 & WFC3/UVIS2 & F336W & 1500 & 2022 Dec 09 & 17181 & MMT/Binospec & $i$\\
SN 2019ujb & 0.1647 & WFC3/UVIS2 & F336W & 1500 & 2023 Feb 4 &  17181 & FLWO/KeplerCam & $i$\\
SN 2019unb & 0.0635 & WFC3/UVIS2 & F275W & 1200 & 2022 Jan 25 & 16657 & P48/MOSAIC & $r$ \\
SN 2019xaq & 0.20 & WFC3/UVIS2 & F336W & 1500 & 2023 Jan 13 & 17181 & P48/MOSAIC & $r$\\
SN 2019zbv & 0.37 & WFC3/UVIS2 & F438W & 1500 & 2022 Dec 18 & 17181 & FLWO/KeplerCam & $i$\\
SN 2019zeu & 0.39 & WFC3/UVIS2 & F438W & 1500 & 2023 Feb 04 & 17181 & FLWO/KeplerCam & $i$\\
SN 2020ank & 0.2485 & WFC3/UVIS2 & F390W & 1500 & 2022 Dec 09 & 17181 & P48/MOSAIC & $r$ \\
SN 2020jhm & 0.05 & WFC3/UVIS2 & F336W & 1500 & 2022 Dec 07 & 17181 & P48/MOSAIC & $i$ \\
SN 2020onb & 0.153 & WFC3/UVIS2 & F336W & 1500 & 2023 Jan 20 & 17181 & P48/MOSAIC & $i$ \\
SN 2020qef & 0.183 & WFC3/UVIS2 & F336W & 1500 & 2023 Jan 04 & 17181 & P48/MOSAIC & $r$ \\
SN 2020qlb & 0.1585 & WFC3/UVIS2 & F336W & 1600 & 2022 Jan 07 & 16657 & P48/MOSAIC & $r$ \\
SN 2020rmv & 0.27 & WFC3/UVIS2 & F336W & 1600 & 2021 Dec 04 & 16657 & P48/MOSAIC & $r$ \\
SN 2020tcw & 0.064 & WFC3/UVIS2 & F275W & 1200 & 2022 Jun 12 & 16657 & P48/MOSAIC & $r$ \\
SN 2020xgd & 0.454 & WFC3/UVIS2 & F438W & 1500 & 2022 Dec 30 & 17181 & P48/MOSAIC & $r$ \\
SN 2020znr & 0.10 & WFC3/UVIS2 & F336W & 1500 & 2022 Dec 09 & 17181 & {\it Magellan}/IMACS & $i$\\
SNLS07D2bv & 1.50 & ACS/WFC & F850LP & 2169 & 2015 Mar 14 & 13641 & \nodata & \nodata \\[+2pt]
\enddata
\tablecomments{List of all SLSNe with archival {\it HST} host galaxy observations. We also report the telescope/instrument and filter for the events with available SN imaging used for astrometric matching.}
\label{tab:objects}
\end{deluxetable*}

We obtained the {\it HST} data from the Mikulski Archive for Space Telescopes (MAST\footnote{https://archive.stsci.edu}), and retrieved charge transfer efficiency (CTE) corrected images with {\tt astroquery}.  The majority of observations were obtained with a standard four-point dither pattern for optimal pixel sub-sampling during a single visit, while some have either a two-point pattern, a three-point pattern, or multiple visits in the same filter. By combining dithered exposures, we reconstruct a final image for each host galaxy with higher resolution than the original ones sampled by the instrumental point-spread function (PSF). We further apply distortion corrections, which improve the precision of our astrometric alignment. The CTE-corrected images were processed and combined with the {\tt AstroDrizzle} task as a part of the {\tt DrizzlePac} software package provided by STScI (\citealt{Gonzaga_2012}). We use {\tt final$\_$pixfrac}=$0.8$ or 0.9 depending on the pixel variation of the output image, and {\tt final$\_$scale}$=0.020\arcsec$ and $0.025\arcsec$ per pixel\footnote{Except for PS1-11aib, for which we keep the original pixel scale of $0.05\arcsec$, for consistency with previous work in \citetalias{Lunnan_2015}.} for WFC3/UVIS and ACS, respectively. All the {\it HST} data used in this paper can be found in MAST: \dataset[10.17909/skn3-8756]{http://dx.doi.org/10.17909/skn3-8756}. 

\subsection{SLSN Imaging}
\label{sec:SN_image}

Precisely locating each SLSN within its host galaxy requires relative astrometry and hence images of the SLSNe.  We use the deepest, highest resolution optical images available. Specifically, we use the PanSTARRS1 Medium Deep Survey (PS1-MDS; \citealt{Chambers_2016}) nightly stacks, Palomar Transient Factory (PTF\footnote{\dataset[10.26131/IRSA155]{https://irsa.ipac.caltech.edu/Missions/ptf.html}}; \citealt{Rau_2009}) and Zwicky Transient Facility (ZTF\footnote{\dataset[10.26131/IRSA539]{https://irsa.ipac.caltech.edu/Missions/ztf.html}}; \citealt{Bellm_2019}) images accessed through the NASA Infrared Processing and Analysis Center Infrared Science Archive\footnote{https://irsa.ipac.caltech.edu/}, Dark Energy Survey (DES; \citealt{DES_2005,DES_2016}) images accessed through NSF's NOIRLab Astro Data  Lab\footnote{https://noirlab.edu/public/projects/astrodatalab/}, Gemini Observatory Archive\footnote{https://archive.gemini.edu/searchform}, and the 2-m Liverpool Telescope (\citealt{Steele_2004}) Data Archive\footnote{https://telescope.livjm.ac.uk}. Finally, in some cases, we use our own data from the 1.2-m telescope at Fred L.~Whipple Observatory (KeplerCam), the 6.5-m {\it Magellan} telescopes (IMACS), and the 6.5-m MMT (Binospec). 

Five SLSNe in our sample have multiple {\it HST} epochs available, which include detections of the SNe themselves. As previously analyzed in \citetalias{Lunnan_2015}, the constant F625W flux and F625W--F775W color in the two final epochs of PS1-11aib (220 and 350 rest-frame days past peak) indicate no residual SN emission in the final epoch. Similarly, SCP06F6 and SN2016eay both have {\it HST} observations taken significantly far apart ($\approx 1170$ and $\approx 1130$ rest-frame days past peak, respectively) to confidently rule out the possibility of residual SN emission in the final epochs. We use the first epoch of {\it HST} observations of PS1-11aib, SCP06F6, and SN2016eay as the SN images, performing image subtraction to ensure that host galaxy light does not affect the SN centroid determination. We use {\tt PyZOGY}\footnote{https://github.com/dguevel/PyZOGY} (\citealt{Zackay_2016}) after aligning the different {\it HST} images with our astrometry procedure (see \S\ref{sec:rel_astrometry}).

On the other hand, for SN\,2015bn and SN\,2016inl, all available {\it HST} observations contain SN emission, and we therefore disentangle the SN and host galaxy contributions using {\tt galfit} (\citealt{Peng_2010}). For SN\,2015bn, we use the Sérsic profile model from \citet{Nicholl_2018} and use the host-subtracted image containing only SN light to calculate the centroid location. In the case of SN\,2016inl, we use the model from \citet{Blanchard_2021}, where the SN and its host galaxy were simultaneously fit with a PSF and a Sérsic profile. Since the {\tt galfit} models do not include positional uncertainties, we take the centroid locations given by {\tt galfit} and measure the SN positional uncertainty in the same fashion as the rest of our sample and assume a typical host positional uncertainty of $\sigma_{\rm host}=0.005''$ for smoothly varying galaxies in the sample (e.g., PS1-10bzj and PS1-11ap). We note that for these 2 SLSNe we can measure an offset, but due to the blending of SN and host galaxy light we cannot reliably determine the fractional flux statistic (see \S\ref{sec:methods}).

Taking into account the availability of SLSN images (ground-based and {\it HST}), the sample size is reduced to 95 SLSNe; we do not find any publicly available SN images for the remaining 14 events; see Table~\ref{tab:seq}.

\subsection{Astrometry}
\label{sec:rel_astrometry}

We perform relative astrometry when possible, and absolute astrometry otherwise, on the 95 {\it HST} images with available SN images, to align and precisely locate each SLSN relative to its host galaxy. 

We identify common point-like sources between the SN and {\it HST} images with {\tt photutils} \citep{Bradley_2022} and measure the position of each source. We then match the sources using the function {\tt wcs.fit\_wcs\_from\_points} in {\tt astropy} to fit a Simple Imaging Polynomial of degree $1-3$ (depending on the number of common sources) to align the SN images to the world coordinate system of the {\it HST} images. We measure the root-mean-square (rms) of the positional residuals for the common sources as the $1\sigma$ astrometric tie uncertainty, $\sigma_{\rm tie}$. We require a minimum of 4 common sources for a reliable fit, which is satisfied for 66 of the 95 SLSNe.  For the remaining 29 events we instead use absolute astrometry by matching both the SN and the {\it HST} images to the Gaia DR3 catalog (\citealt{Gaia_2016,Gaia_2023}). The rms positional residuals from each fit are then combined in quadrature to determine $\sigma_{\rm tie}$.

For the sources aligned with relative astrometry, we find a median value of $\sigma_{\rm tie}\approx 33$ mas. For sources  aligned with absolute astrometry, the combined tie uncertainty is somewhat higher, with a median value of $\sigma_{\rm tie}\approx 87$ mas. The number of common sources used in the astrometric matching, and the resulting values of $\sigma_{\rm tie}$ are listed in Table~\ref{tab:astrometry}. We note that for 11 sources, there are a lack of point-like sources in the {\it HST} images, rendering astrometric match impossible. The sample size after astrometric alignment is reduced to 84 SLSNe; see Table~\ref{tab:seq}.

\begin{figure*}[t!]
    \figurenum{2}
    \centering
    \includegraphics[width=0.95\textwidth]{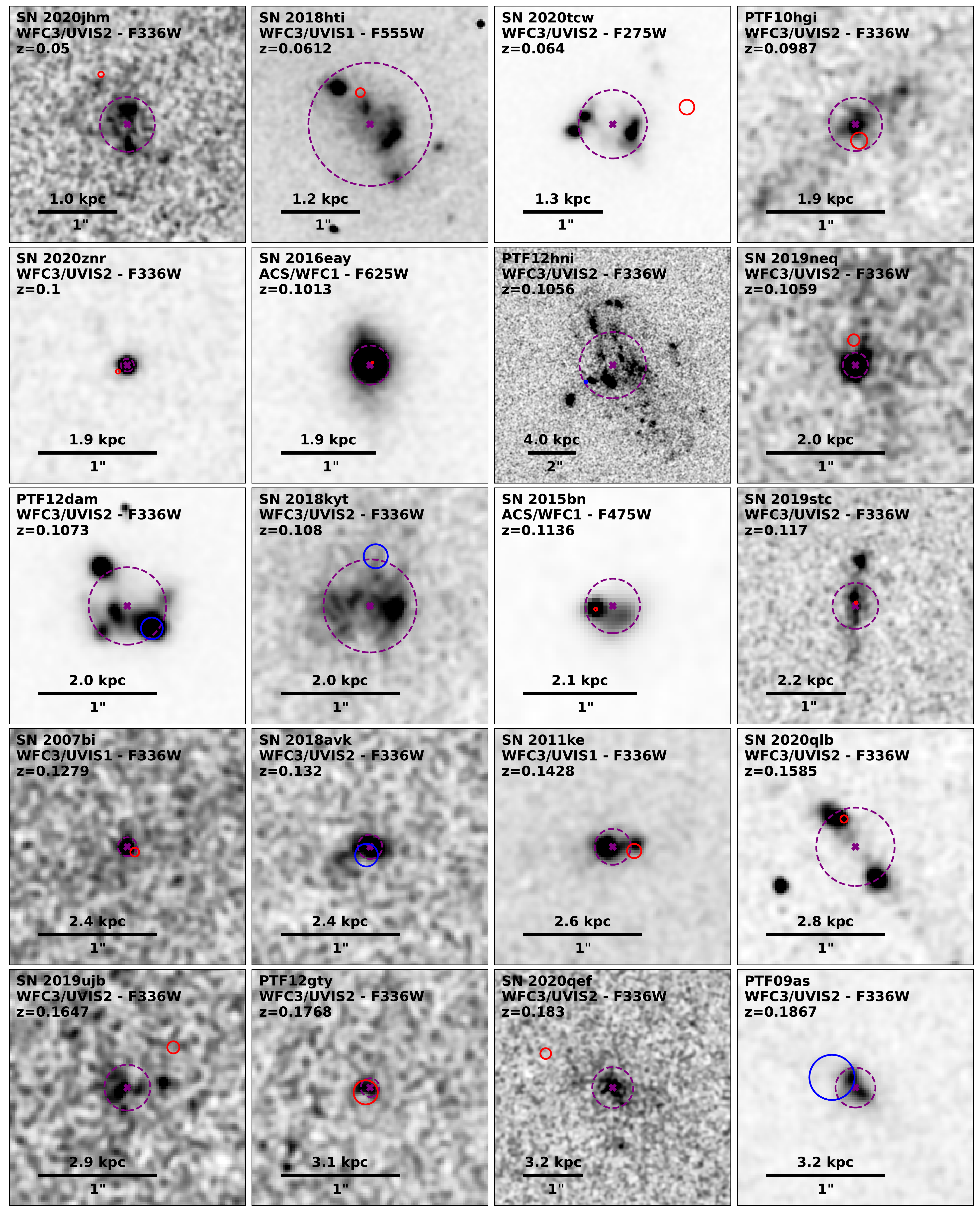}
    \caption{{\it HST} images of the 65 SLSN host galaxies with available SN imaging and successful astrometric alignment ({\it HST} images of non-detected hosts are shown in Figure~\ref{fig:offsets_nogal}). The images are centered on the centroid of each host galaxy (purple crosses) and aligned with North up and East to the left. The dashed purple circles marks $R_{50}$ (half-light radius). Solid circles mark the location of the SLSNe, with a radius corresponding to the $1\sigma$ uncertainty. Red and blue circles indicate positions determined using relative or absolute astrometry, respectively. The images have been smoothed with a 3 pixel Gaussian filter. In the case of SN\,2015bn and SN\,2016inl, the images also contain light from the SLSNe.}
    \label{fig:offsets}
\end{figure*}

\begin{figure*}
    \figurenum{2}
    \centering
    \includegraphics[width=\textwidth]{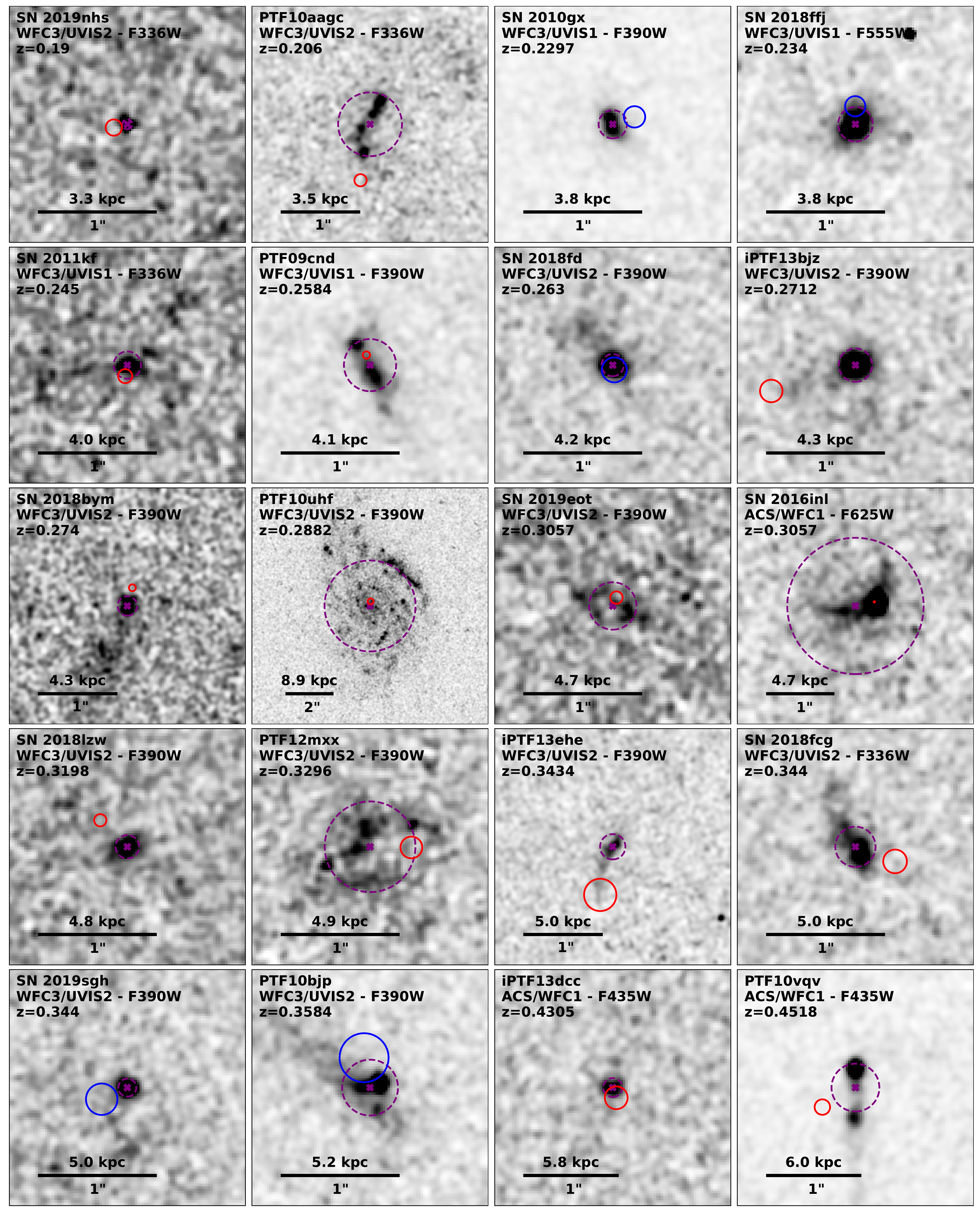}
    \caption{({\it continued})}
\end{figure*}

\begin{figure*}
    \figurenum{2}
    \centering
    \includegraphics[width=\textwidth]{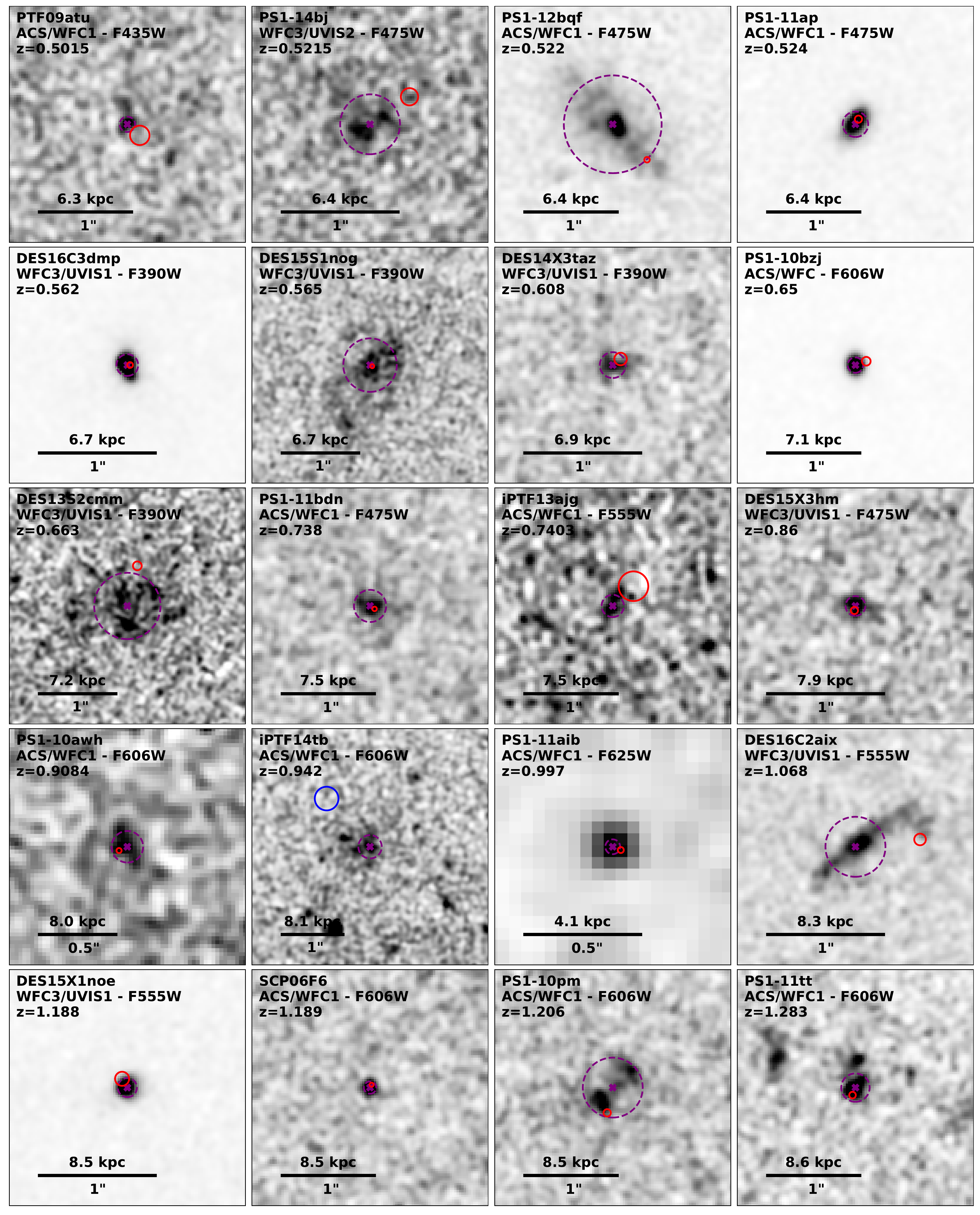}
    \caption{({\it continued})}
\end{figure*}

\begin{figure*}[t!]
    \figurenum{2}
    \centering
    \includegraphics[width=\textwidth]{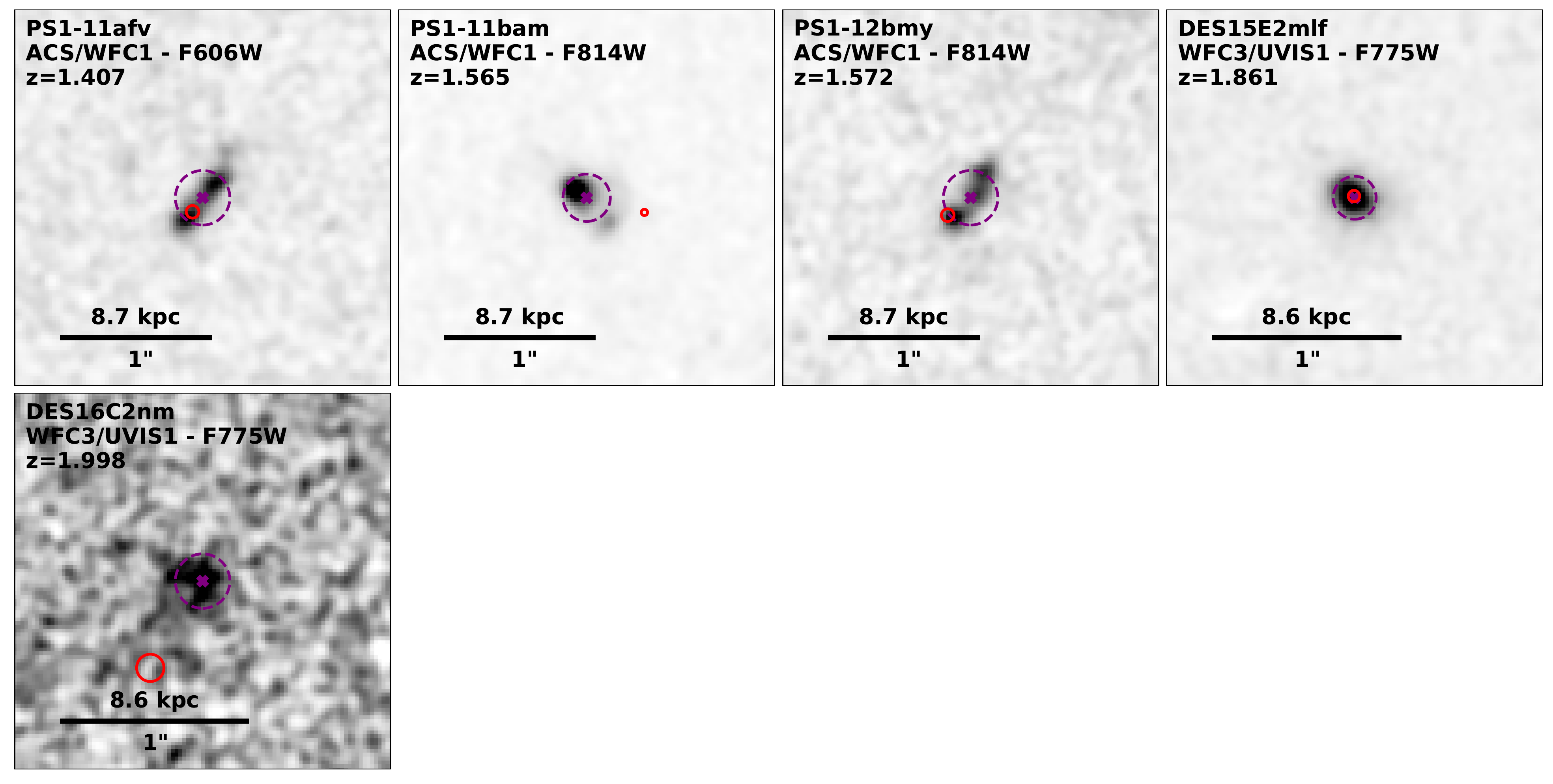}
    \caption{({\it continued})}
\end{figure*}

\section{Measurement Methodology}
\label{sec:methods}

\subsection{Host Galaxy Assignment}
\label{sec:host_assignment}

Before offsets and fractional flux values can be determined, we need to associate a host galaxy to each SLSN. We assign the most probable host galaxy for each SLSN by calculating the probability of chance coincidence ($P_{\rm cc}$) for galaxies in the vicinity of the SN location, specifically focusing on a region of $30\times 30$ kpc. For context, the largest projected physical offsets measured in \citetalias{Lunnan_2015} was 4.3 kpc, and the offsets measured in this work are well-contained within this field of view.

Following the methodology of \citet{Bloom_2002} (as also used by \citealt{Berger_2010}, \citealt{Fong_2013}, and \citetalias{Blanchard_2016}), we calculate $P_{\rm cc}$ for each galaxy using: 
\begin{equation}
\label{eq:p_cc}
P_{\rm cc} = 1 - e^{-\pi {R_e}^2\sigma(\le m)},
\end{equation}
where $\sigma(\leq m)$ is the observed surface density of galaxies brighter than magnitude $m$ and $R_e$ is the effective radius, defined as $R_e=\max(3\sqrt{\sigma_{\rm tie}^2+\sigma_{\rm SN}^2},\sqrt{R^2+4R_{50}^2})$ \citep{Bloom_2002}, where $R$ is the projected offset and $R_{50}$ is the half-light radius (see \S\ref{sec:offsets}). For about $2/3$ of our sample, the localizations are sufficiently precise such that the second term dominates. 

For each {\it HST} image, we use {\tt photutils} to detect extended sources that are potential host candidates. To avoid spurious associations with noise fluctuations, we consider sources with a minimum of 10 connected pixels detected at $\ge 2.5\sigma$ above the background level. If the object with the lowest $P_{\rm cc}$ exceeds a threshold value of 0.1, we consider the actual host galaxy to be undetected at the limit of our images, and exclude the source from subsequent analysis; this is the case for 19 sources of the 84 sources with successful astrometric matching.

In addition to calculating $P_{\rm cc}$, we also compare our most probable host galaxy assignment with previous studies that have  identified SLSN hosts (\citetalias{Lunnan_2015}; \citealt{Angus_2016,Angus_2019,Perley_2016,Cikota_2017,Schulze_2018,Taggart_2021}). We find that our assignments are in excellent agreement with previous studies, including those with the most complex morphology (e.g., PTF12dam, PTF12hni).  In 7 cases\footnote{These objects are DES13S2cmm, PS1-10pm, PS1-14bj, PTF10uhf, PTF12hni, SN2019eot, and SN2019ujb.}, we find multiple objects in the vicinity of the SLSN position, which we consider to be disjointed components of the same host galaxy (due to patchy nature of UV emission). In these instances, we combine the components to calculate all relevant host properties. For the remainder of the sample, the most probably host galaxy is either coincident with the SLSN position or the most proximal extended source.

Our final sample of detected host galaxies contains 65 events (Table~\ref{tab:seq}), which is nearly four times larger than the sample analyzed in \cite{Lunnan_2015}. The assigned host galaxies have $P_{\rm cc}\approx 1.8\times 10^{-4} - 7.5\times 10^{-2}$; see Table~\ref{tab:astrometry}. In Figure~\ref{fig:offsets} we show the drizzled {\it HST} images, along with the location and uncertainty region of each SLSN (the quadrature sum of $\sigma_{\rm tie}$ and the SN centroid uncertainty, $\sigma_{\rm SN}$; \S\ref{sec:offsets}), as well as the centroid and the half-light radius of each host galaxy (see \S\ref{sec:offsets}).  The drizzled {\it HST} images for SLSNe without an identified host are shown in Figure~\ref{fig:offsets_nogal}.

\subsection{Offset Measurements}
\label{sec:offsets}

\begin{deluxetable}{lc}
\tablewidth{\columnwidth} 
\caption{SLSN Sample with {\it HST} Data \label{tab:seq}}
\tablehead{\colhead{Criterion} & \colhead{Number of Events}}
\startdata
Preliminary Sample & 109\\
Available SN Image & 95\\
Successful Astrometric Match & 84\\
Host Galaxy Detected & 65\\[+2pt]
\enddata
\end{deluxetable}

Following the astrometric matching and host galaxy assignment, we determine the angular offset of each SLSN from the UV light centroid of its host galaxy. We determine the location of each SLSN by fitting the SN image with a 2D Gaussian and calculate the image centroid and its associated uncertainty, $\sigma_{\rm SN}$, which is determined as $\theta_{\rm FWHM}/{\rm 2(S/N)}$, where $\theta_{\rm FWHM}$ is the full-width at half maximum of the 2D Gaussian, and ${\rm S/N}$ is the signal-to-noise ratio of the SN detection; see Table~\ref{tab:astrometry}. 

Next, we define the host galaxy flux-weighted centroid, as determined by {\tt photutils} for pixels designated as part of each host galaxy. Here, the statistical uncertainty on the host centroid from {\tt photutils} potentially underestimates the systematic uncertainties due to the irregular morphology of some hosts. To address this, we estimate the positional uncertainty ($\sigma_{\rm host}$) by varying the detection threshold from $2.5\sigma$ to the highest threshold at which the host is still detected with a step size of $0.005\sigma$, and then taking the standard deviation of the resulting host centroid values. The resulting values of $\sigma_{\rm host}$ are listed in Table~\ref{tab:astrometry}.

For each SLSN we measure the projected {\it physical} offset, $R_{\rm phys}$, using the SN redshift, and assign an associated total uncertainty of $\sigma_{R_{\rm phys}}=\sqrt{\sigma_{\rm tie}^2+\sigma_{\rm SN}^2+\sigma_{\rm host}^2}$. In addition to the projected physical offset, we also normalize the offset of each SLSN by the size of its host galaxy, $R_{\rm norm}$, allowing us to explore both the population itself and to compare it to other transients that may arise in galaxies with different sizes.  We use {\tt photutils} to measure the half-light radius, $R_{50}$, defined as the effective circular radius that encloses 50\% of the total flux within the galaxy Kron aperture. Using $R_{50}$, the host-normalized offsets is simply given by $R_{\rm norm}=R_{\rm phys}/R_{50}$ (Table~\ref{tab:astrometry}).

\subsection{Fractional Flux Measurements}
\label{sec:frac_flux_measurement}

\startlongtable
\begin{deluxetable*}{lccccccccccc}
\tablewidth{\textwidth} 
\tablecaption{Astrometric Results and Key Measurements}
\tablehead{
\colhead{} & \colhead{} & \colhead{$\sigma_{\rm tie}$} & \colhead{$\sigma_{\rm SN}$} & \colhead{$\sigma_{\rm host}$} & \colhead{$R_{\rm phy}$} & \colhead{$R_{50}$} & \colhead{$R_{\rm norm}$} & \colhead{Light} & \colhead{} & \colhead{}\\[-12pt]
\colhead{SLSN} & \colhead{\# Tie Objects} & \colhead{} & \colhead{} & \colhead{} & \colhead{} & \colhead{} & \colhead{} & \colhead{} & \colhead{AB Mag\tablenotemark{$\rm *$}} & \colhead{$P_{\rm cc}$} \\[-13pt]
\colhead{} & \colhead{} & \colhead{(")} & \colhead{(")} & \colhead{(")} & \colhead{(kpc)} & \colhead{(kpc)} & \colhead{$(R_{\rm phy}/R_{50})$} & \colhead{Fraction} & \colhead{} & \colhead{}}
\startdata
SN 2020jhm & 4 & 0.033 & 0.011 & \phn0.008 & $0.724^{+0.037}_{-0.037}$ & 0.352 & $2.056^{+0.104}_{-0.104}$ & $0.000^{+0.000}_{-0.000}$ & $22.632^{+0.036}_{-0.036}$ & \phn0.006 \\
SN 2018hti & 6 & 0.055 & 0.013 & \phn0.010 & $0.513^{+0.070}_{-0.070}$ & 0.956 & $0.537^{+0.074}_{-0.074}$ & $0.599^{+0.075}_{-0.331}$ & $19.063^{+0.006}_{-0.006}$ & \phn0.002 \\
SN 2020tcw & 5 & 0.094 & 0.010 & \phn0.007 & $1.230^{+0.120}_{-0.120}$ & 0.553 & $2.224^{+0.218}_{-0.218}$ & $0.000^{+0.023}_{-0.000}$ & $19.206^{+0.004}_{-0.004}$ & \phn0.001 \\
PTF10hgi & 8 & 0.067 & 0.016 & \phn0.011 & $0.268^{+0.131}_{-0.131}$ & 0.426 & $0.628^{+0.308}_{-0.308}$ & $0.250^{+0.464}_{-0.250}$ & $24.398^{+0.028}_{-0.028}$ & \phn0.006 \\
SN 2020znr & 24 & 0.018 & 0.001 & \phn0.002 & $0.183^{+0.035}_{-0.035}$ & 0.104 & $1.766^{+0.335}_{-0.335}$ & $0.222^{+0.129}_{-0.183}$ & $23.042^{+0.016}_{-0.016}$ & $<$0.001 \\
SN 2016eay & 13 & 0.005 & 0.008 & \phn0.008 & $0.072^{+0.024}_{-0.024}$ & 0.399 & $0.180^{+0.059}_{-0.059}$ & $0.924^{+0.067}_{-0.012}$ & $21.983^{+0.005}_{-0.005}$ & \phn0.001 \\
PTF12hni\tablenotemark{$\dagger$} & 12,5 & 0.047 & 0.032 & $<$0.001 & $2.667^{+0.114}_{-0.114}$ & 2.811 & $0.949^{+0.041}_{-0.041}$ & $0.000^{+0.205}_{-0.000}$ & $20.320^{+0.011}_{-0.011}$ & \phn0.010 \\
SN 2019neq & 5 & 0.047 & 0.014 & \phn0.007 & $0.426^{+0.099}_{-0.099}$ & 0.213 & $2.005^{+0.468}_{-0.468}$ & $0.034^{+0.289}_{-0.034}$ & $23.170^{+0.024}_{-0.024}$ & \phn0.001 \\
PTF12dam\tablenotemark{$\dagger$} & 8,4 & 0.091 & 0.008 & \phn0.008 & $0.571^{+0.187}_{-0.187}$ & 0.665 & $0.858^{+0.281}_{-0.281}$ & $0.900^{+0.063}_{-0.116}$ & $19.977^{+0.007}_{-0.007}$ & $<$0.001 \\
SN 2018kyt\tablenotemark{$\dagger$} & 8,3 & 0.093 & 0.041 & \phn0.012 & $0.864^{+0.209}_{-0.209}$ & 0.804 & $1.075^{+0.260}_{-0.260}$ & $0.180^{+0.319}_{-0.180}$ & $21.644^{+0.015}_{-0.015}$ & \phn0.002 \\
SN 2015bn\tablenotemark{$\ddagger$} & \nodata & \nodata & 0.014 & \phn0.005 & $0.462^{+0.031}_{-0.031}$ & 0.517 & $0.894^{+0.061}_{-0.061}$ & \nodata & \nodata & \nodata \\
SN 2019stc & 52 & 0.016 & 0.006 & \phn0.007 & $0.093^{+0.040}_{-0.040}$ & 0.637 & $0.146^{+0.063}_{-0.063}$ & $0.808^{+0.052}_{-0.232}$ & $22.637^{+0.029}_{-0.029}$ & \phn0.002 \\
SN 2007bi & 7 & 0.037 & 0.008 & \phn0.003 & $0.183^{+0.089}_{-0.089}$ & 0.189 & $0.971^{+0.469}_{-0.469}$ & $0.144^{+0.566}_{-0.144}$ & $25.245^{+0.069}_{-0.069}$ & \phn0.001 \\
SN 2018avk\tablenotemark{$\dagger$} & 6,3 & 0.052 & 0.081 & \phn0.007 & $0.187^{+0.234}_{-0.187}$ & 0.249 & $0.748^{+0.940}_{-0.748}$ & $0.418^{+0.344}_{-0.418}$ & $24.117^{+0.041}_{-0.041}$ & \phn0.001 \\
SN 2011ke & 4 & 0.059 & 0.012 & \phn0.004 & $0.481^{+0.157}_{-0.157}$ & 0.395 & $1.216^{+0.398}_{-0.398}$ & $0.431^{+0.214}_{-0.248}$ & $22.861^{+0.016}_{-0.016}$ & \phn0.001 \\
SN 2020qlb & 10 & 0.028 & 0.013 & $<$0.001 & $0.719^{+0.087}_{-0.087}$ & 0.938 & $0.767^{+0.093}_{-0.093}$ & $0.780^{+0.174}_{-0.190}$ & $22.642^{+0.019}_{-0.019}$ & \phn0.003 \\
SN 2019ujb & 6 & 0.037 & 0.036 & \phn0.008 & $1.510^{+0.152}_{-0.152}$ & 0.566 & $2.667^{+0.268}_{-0.268}$ & $0.000^{+0.212}_{-0.000}$ & $24.048^{+0.080}_{-0.080}$ & \phn0.007 \\
PTF12gty & 6 & 0.039 & 0.095 & \phn0.005 & $0.169^{+0.317}_{-0.169}$ & 0.243 & $0.693^{+1.301}_{-0.693}$ & $0.394^{+0.399}_{-0.394}$ & $26.030^{+0.098}_{-0.098}$ & \phn0.008 \\
SN 2020qef & 12 & 0.043 & 0.080 & \phn0.015 & $4.044^{+0.292}_{-0.292}$ & 1.102 & $3.670^{+0.265}_{-0.265}$ & $0.000^{+0.000}_{-0.000}$ & $22.575^{+0.025}_{-0.025}$ & \phn0.012 \\
PTF09as\tablenotemark{$\dagger$} & 12,4 & 0.148 & 0.122 & \phn0.006 & $0.700^{+0.618}_{-0.618}$ & 0.546 & $1.282^{+1.132}_{-1.132}$ & $0.164^{+0.271}_{-0.164}$ & $23.269^{+0.018}_{-0.018}$ & \phn0.003 \\
SN 2019nhs & 4 & 0.054 & 0.042 & \phn0.004 & $0.386^{+0.226}_{-0.226}$ & 0.139 & $2.785^{+1.626}_{-1.626}$ & $0.000^{+0.012}_{-0.000}$ & $25.664^{+0.083}_{-0.083}$ & \phn0.003 \\
PTF10aagc & 4 & 0.047 & 0.062 & \phn0.005 & $2.515^{+0.271}_{-0.271}$ & 1.415 & $1.778^{+0.192}_{-0.192}$ & $0.000^{+0.053}_{-0.000}$ & $22.980^{+0.029}_{-0.029}$ & \phn0.009 \\
SN 2010gx\tablenotemark{$\dagger$} & 5,8 & 0.090 & 0.001 & \phn0.008 & $0.738^{+0.344}_{-0.344}$ & 0.455 & $1.623^{+0.757}_{-0.757}$ & $0.004^{+0.231}_{-0.004}$ & $23.823^{+0.021}_{-0.021}$ & \phn0.001 \\
SN 2018ffj\tablenotemark{$\dagger$} & 8,5 & 0.071 & 0.048 & \phn0.001 & $0.586^{+0.329}_{-0.329}$ & 0.562 & $1.043^{+0.585}_{-0.585}$ & $0.444^{+0.116}_{-0.250}$ & $23.343^{+0.020}_{-0.020}$ & \phn0.001 \\
SN 2011kf & 4 & 0.022 & 0.055 & \phn0.006 & $0.378^{+0.238}_{-0.238}$ & 0.460 & $0.823^{+0.519}_{-0.519}$ & $0.000^{+0.356}_{-0.000}$ & $25.197^{+0.067}_{-0.067}$ & \phn0.002 \\
PTF09cnd & 5 & 0.012 & 0.029 & \phn0.009 & $0.375^{+0.134}_{-0.134}$ & 0.914 & $0.410^{+0.146}_{-0.146}$ & $0.558^{+0.112}_{-0.186}$ & $23.915^{+0.023}_{-0.023}$ & \phn0.003 \\
SN 2018fd\tablenotemark{$\dagger$} & 8,10 & 0.084 & 0.064 & \phn0.005 & $0.179^{+0.443}_{-0.179}$ & 0.393 & $0.454^{+1.128}_{-0.454}$ & $0.712^{+0.192}_{-0.237}$ & $24.216^{+0.025}_{-0.025}$ & \phn0.002 \\
iPTF13bjz & 4 & 0.036 & 0.088 & \phn0.012 & $3.194^{+0.412}_{-0.412}$ & 0.604 & $5.285^{+0.681}_{-0.681}$ & $0.000^{+0.012}_{-0.000}$ & $24.608^{+0.028}_{-0.028}$ & \phn0.016 \\
SN 2018bym & 17 & 0.034 & 0.027 & \phn0.007 & $1.035^{+0.191}_{-0.191}$ & 0.531 & $1.949^{+0.360}_{-0.360}$ & $0.000^{+0.000}_{-0.000}$ & $24.933^{+0.055}_{-0.055}$ & \phn0.004 \\
PTF10uhf & 6 & 0.097 & 0.080 & \phn0.008 & $0.876^{+0.564}_{-0.564}$ & 8.613 & $0.102^{+0.066}_{-0.066}$ & $0.677^{+0.230}_{-0.269}$ & $20.180^{+0.006}_{-0.006}$ & \phn0.012 \\
SN 2019eot & 10 & 0.026 & 0.046 & \phn0.009 & $0.361^{+0.249}_{-0.249}$ & 0.934 & $0.386^{+0.266}_{-0.266}$ & $0.000^{+0.444}_{-0.000}$ & $24.439^{+0.111}_{-0.111}$ & \phn0.004 \\
SN 2016inl\tablenotemark{$\ddagger$} & \nodata & \nodata & 0.010 & \phn0.005 & $1.335^{+0.051}_{-0.051}$ & 4.705 & $0.284^{+0.011}_{-0.011}$ & \nodata & \nodata & \nodata \\
SN 2018lzw & 8 & 0.035 & 0.039 & \phn0.008 & $1.540^{+0.254}_{-0.254}$ & 0.477 & $3.232^{+0.533}_{-0.533}$ & $0.000^{+0.044}_{-0.000}$ & $25.292^{+0.058}_{-0.058}$ & \phn0.006 \\
PTF12mxx & 10 & 0.039 & 0.083 & \phn0.005 & $1.715^{+0.452}_{-0.452}$ & 1.882 & $0.911^{+0.240}_{-0.240}$ & $0.000^{+0.097}_{-0.000}$ & $24.951^{+0.038}_{-0.038}$ & \phn0.024 \\
iPTF13ehe & 10 & 0.073 & 0.192 & \phn0.013 & $3.179^{+1.035}_{-1.035}$ & 0.810 & $3.926^{+1.279}_{-1.279}$ & $0.000^{+0.244}_{-0.000}$ & $25.235^{+0.042}_{-0.042}$ & \phn0.022 \\
SN 2018fcg & 29 & 0.087 & 0.050 & \phn0.026 & $1.803^{+0.522}_{-0.522}$ & 0.859 & $2.099^{+0.608}_{-0.608}$ & $0.000^{+0.009}_{-0.000}$ & $22.720^{+0.032}_{-0.032}$ & \phn0.002 \\
SN 2019sgh\tablenotemark{$\dagger$} & 13,13 & 0.075 & 0.110 & \phn0.005 & $1.203^{+0.671}_{-0.671}$ & 0.383 & $3.142^{+1.753}_{-1.753}$ & $0.000^{+0.160}_{-0.000}$ & $24.960^{+0.046}_{-0.046}$ & \phn0.006 \\
PTF10bjp\tablenotemark{$\dagger$} & 20,4 & 0.150 & 0.143 & \phn0.003 & $1.336^{+1.073}_{-1.073}$ & 1.224 & $1.091^{+0.877}_{-0.877}$ & $0.000^{+0.272}_{-0.000}$ & $23.835^{+0.021}_{-0.021}$ & \phn0.006 \\
iPTF13dcc & 4 & 0.057 & 0.106 & \phn0.011 & $0.661^{+0.700}_{-0.700}$ & 0.582 & $1.135^{+1.203}_{-1.203}$ & $0.007^{+0.519}_{-0.007}$ & $26.018^{+0.066}_{-0.066}$ & \phn0.011 \\
PTF10vqv & 5 & 0.027 & 0.077 & \phn0.011 & $2.419^{+0.492}_{-0.492}$ & 1.507 & $1.605^{+0.326}_{-0.326}$ & $0.000^{+0.000}_{-0.000}$ & $23.804^{+0.018}_{-0.018}$ & \phn0.006 \\
PTF09atu & 10 & 0.069 & 0.076 & \phn0.007 & $1.111^{+0.650}_{-0.650}$ & 0.512 & $2.170^{+1.269}_{-1.269}$ & $0.000^{+0.154}_{-0.000}$ & $27.566^{+0.104}_{-0.104}$ & \phn0.024 \\
PS1-14bj & 6 & 0.052 & 0.052 & \phn0.012 & $2.620^{+0.477}_{-0.477}$ & 1.630 & $1.607^{+0.293}_{-0.293}$ & $0.049^{+0.547}_{-0.049}$ & $25.012^{+0.080}_{-0.080}$ & \phn0.014 \\
PS1-12bqf & 18 & 0.024 & 0.017 & \phn0.022 & $3.359^{+0.235}_{-0.235}$ & 3.330 & $1.009^{+0.071}_{-0.071}$ & $0.650^{+0.114}_{-0.300}$ & $22.946^{+0.012}_{-0.012}$ & \phn0.009 \\
PS1-11ap & 14 & 0.038 & 0.004 & \phn0.007 & $0.403^{+0.251}_{-0.251}$ & 0.869 & $0.464^{+0.288}_{-0.288}$ & $0.899^{+0.057}_{-0.133}$ & $24.151^{+0.011}_{-0.011}$ & \phn0.001 \\
DES16C3dmp & 15 & 0.017 & 0.017 & \phn0.002 & $0.154^{+0.164}_{-0.154}$ & 0.612 & $0.252^{+0.267}_{-0.252}$ & $0.764^{+0.105}_{-0.164}$ & $22.963^{+0.008}_{-0.008}$ & $<$0.001 \\
DES15S1nog & 12 & 0.020 & 0.016 & \phn0.005 & $0.208^{+0.173}_{-0.173}$ & 2.275 & $0.092^{+0.076}_{-0.076}$ & $0.544^{+0.343}_{-0.350}$ & $23.862^{+0.027}_{-0.027}$ & \phn0.007 \\
DES14X3taz & 6 & 0.015 & 0.051 & \phn0.010 & $0.583^{+0.375}_{-0.375}$ & 0.759 & $0.767^{+0.494}_{-0.494}$ & $0.099^{+0.483}_{-0.099}$ & $25.882^{+0.068}_{-0.068}$ & \phn0.004 \\
PS1-10bzj & 28 & 0.040 & 0.023 & \phn0.001 & $0.867^{+0.329}_{-0.329}$ & 0.581 & $1.491^{+0.567}_{-0.567}$ & $0.292^{+0.130}_{-0.097}$ & $24.157^{+0.006}_{-0.006}$ & \phn0.001 \\
DES13S2cmm & 19 & 0.047 & 0.030 & \phn0.005 & $3.778^{+0.404}_{-0.404}$ & 3.031 & $1.246^{+0.133}_{-0.133}$ & $0.000^{+0.000}_{-0.000}$ & $23.897^{+0.044}_{-0.044}$ & \phn0.015 \\
PS1-11bdn & 5 & 0.023 & 0.007 & \phn0.003 & $0.428^{+0.183}_{-0.183}$ & 1.278 & $0.335^{+0.143}_{-0.143}$ & $0.824^{+0.075}_{-0.354}$ & $25.971^{+0.058}_{-0.058}$ & \phn0.008 \\
iPTF13ajg & 6 & 0.059 & 0.145 & \phn0.018 & $2.274^{+1.187}_{-1.187}$ & 0.884 & $2.571^{+1.342}_{-1.342}$ & $0.000^{+0.000}_{-0.000}$ & $27.515^{+0.113}_{-0.113}$ & \phn0.049 \\
DES15X3hm & 18 & 0.029 & 0.010 & \phn0.006 & $0.309^{+0.245}_{-0.245}$ & 0.660 & $0.468^{+0.372}_{-0.372}$ & $0.616^{+0.250}_{-0.428}$ & $27.027^{+0.066}_{-0.066}$ & \phn0.005 \\
PS1-10awh & 22 & 0.009 & 0.014 & \phn0.014 & $0.465^{+0.171}_{-0.171}$ & 0.804 & $0.578^{+0.213}_{-0.213}$ & $0.590^{+0.317}_{-0.496}$ & $26.870^{+0.120}_{-0.120}$ & \phn0.007 \\
iPTF14tb\tablenotemark{$\dagger$} & 24,6 & 0.166 & 0.088 & \phn0.016 & $6.686^{+1.529}_{-1.529}$ & 1.506 & $4.439^{+1.015}_{-1.015}$ & $0.000^{+0.000}_{-0.000}$ & $26.410^{+0.044}_{-0.044}$ & \phn0.076 \\
PS1-11aib & 28 & 0.004 & 0.012 & $<$0.001 & $0.302^{+0.106}_{-0.106}$ & 0.259 & $1.169^{+0.408}_{-0.408}$ & $0.605^{+0.203}_{-0.327}$ & $26.587^{+0.123}_{-0.123}$ & \phn0.001 \\
DES16C2aix & 17 & 0.030 & 0.039 & \phn0.005 & $4.597^{+0.414}_{-0.414}$ & 2.122 & $2.166^{+0.195}_{-0.195}$ & $0.000^{+0.374}_{-0.000}$ & $24.665^{+0.027}_{-0.027}$ & \phn0.014 \\
DES15X1noe & 11 & 0.012 & 0.059 & \phn0.006 & $0.740^{+0.511}_{-0.511}$ & 0.688 & $1.075^{+0.742}_{-0.742}$ & $0.439^{+0.279}_{-0.242}$ & $24.304^{+0.014}_{-0.014}$ & \phn0.001 \\
SCP06F6 & 17 & 0.011 & 0.019 & \phn0.002 & $0.291^{+0.187}_{-0.187}$ & 0.557 & $0.523^{+0.336}_{-0.336}$ & $0.802^{+0.136}_{-0.288}$ & $28.254^{+0.085}_{-0.085}$ & \phn0.007 \\
PS1-10pm & 12 & 0.023 & 0.032 & \phn0.003 & $2.333^{+0.333}_{-0.333}$ & 2.703 & $0.863^{+0.123}_{-0.123}$ & $0.293^{+0.371}_{-0.249}$ & $25.169^{+0.046}_{-0.046}$ & \phn0.018 \\
PS1-11tt & 16 & 0.028 & 0.021 & \phn0.002 & $0.734^{+0.305}_{-0.305}$ & 1.290 & $0.569^{+0.237}_{-0.237}$ & $0.672^{+0.186}_{-0.384}$ & $26.025^{+0.043}_{-0.043}$ & \phn0.007 \\
PS1-11afv & 20 & 0.028 & 0.032 & \phn0.006 & $0.994^{+0.368}_{-0.368}$ & 1.570 & $0.633^{+0.235}_{-0.235}$ & $0.761^{+0.131}_{-0.215}$ & $25.465^{+0.028}_{-0.028}$ & \phn0.007 \\
PS1-11bam & 18 & 0.018 & 0.011 & \phn0.007 & $3.445^{+0.194}_{-0.194}$ & 1.368 & $2.519^{+0.142}_{-0.142}$ & $0.020^{+0.122}_{-0.020}$ & $23.942^{+0.014}_{-0.014}$ & \phn0.004 \\
PS1-12bmy & 10 & 0.011 & 0.040 & \phn0.004 & $1.662^{+0.367}_{-0.367}$ & 1.574 & $1.056^{+0.233}_{-0.233}$ & $0.720^{+0.223}_{-0.575}$ & $25.078^{+0.032}_{-0.032}$ & \phn0.006 \\
DES15E2mlf & 13 & 0.019 & 0.024 & \phn0.004 & $0.085^{+0.267}_{-0.085}$ & 0.988 & $0.086^{+0.270}_{-0.086}$ & $0.910^{+0.052}_{-0.094}$ & $23.464^{+0.014}_{-0.014}$ & \phn0.001 \\
DES16C2nm & 4 & 0.012 & 0.072 & \phn0.002 & $4.624^{+0.623}_{-0.623}$ & 1.243 & $3.720^{+0.501}_{-0.501}$ & $0.000^{+0.132}_{-0.000}$ & $25.283^{+0.067}_{-0.067}$ & \phn0.015 \\[+2pt]
\enddata
\tablenotetext{$\rm *$}{Corrected for Galactic extinction.}
\tablenotetext{\dagger}{Aligned with absolute astrometry using Gaia DR3 catalog.  In these cases we report the number of common sources between the Gaia DR3 catalog and the SN and {\it HST} images, respectively, in the second column.}
\tablenotetext{\ddagger}{The {\it HST} images for these sources contain residual SN emission, allowing for direct astrometry without a tie to another image, but preventing a determination of fractional flux.}

\label{tab:astrometry}
\end{deluxetable*}

As in previous studies of the locations of transients in their host galaxies (\citealt{Fruchter_2006,Kelly_2008,Prieto_2008,Kelly_Kirshner_2012}; \citetalias{Lunnan_2015}; \citetalias{Blanchard_2016}), we measure the fractional flux statistic for each SLSN following the methodology of \citet{Fruchter_2006}, with a refined procedure to assess uncertainties. The fractional flux quantifies the fraction of total flux from the host galaxy that is contained in pixels {\it fainter} than the flux value at the SLSN location, thereby providing a statistic that measures the brightness of the SLSN location compared to the entire galaxy. The resulting fractional flux value lies between zero and one, where a value of one indicates the SLSN occurred in the brightest region of its host galaxy.

\begin{figure*}[t!]
    \centering
    \includegraphics[width=0.9\textwidth]{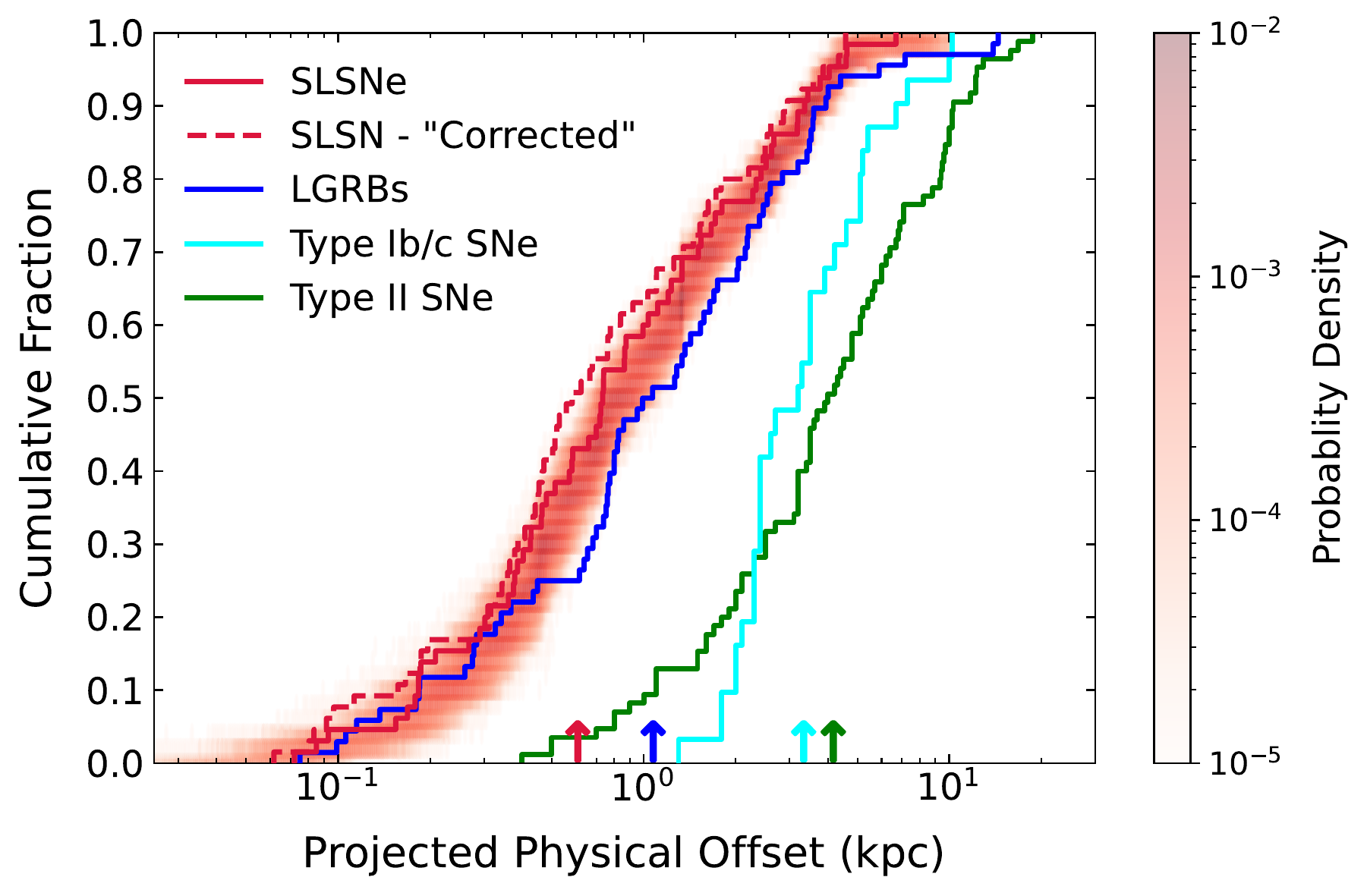}
    \caption{Cumulative distribution of projected physical offsets for 65 SLSNe from this work (red). Also shown are the distributions for LGRBs (blue; \citetalias{Blanchard_2016}), SNe Ib/c (cyan) and II (green) from \citet{Prieto_2008}. The red shaded region shows the results of our Monte Carlo simulation using the associated uncertainties as a 2D histogram. Arrows at the bottom denote the medians of each distribution. The dashed red line includes a correction factor to account for the positive-definite nature of offsets (see Appendix~\ref{sec:correction} for details).}
    \label{fig:offsets_mc}
\end{figure*}

In previous studies, the fractional flux was measured using an area-averaged flux value centered at the transient position, with an error circle radius given by the quadrature sum of $\sigma_{\rm tie}$ and $\sigma_{\rm trans}$. Here, we refine this procedure using a Monte Carlo approach as follows. First, we use {\tt photutils} to extract host galaxy pixels, using a minimum of 10 connected pixels and a threshold of $1\sigma$ above the sky background. Second, exploiting the quasi-Gaussian nature of $\sigma_{\rm tie}$ and $\sigma_{\rm SN}$, we construct a 2D Gaussian probability distribution function centered at the pixel $(x_{\rm SN},y_{\rm SN})$ corresponding to the SLSN centroid, with a standard deviation of $\sigma=\sqrt{\sigma_{\rm tie}^2+\sigma_{\rm SN}^2}$. Third, we randomly sample a pixel based on the 2D Gaussian probability distribution associated with this pixel, and use the flux value at the chosen pixel to calculate the fractional flux by dividing the sum of flux values in pixels dimmer than the chosen pixel by the total flux of the host. To properly determine the uncertainty on the fractional flux, we repeat this procedure 10,000 times. The resulting median values, as well as the $1\sigma$ ranges (corresponding to 16th and 84th percentiles), are reported in Table~\ref{tab:astrometry}.

\section{Results}
\label{sec:results}

Our final sample of 65 SLSNe with offset and fractional flux measurements is four times larger than in the previous study of \citetalias{Lunnan_2015}. In this section we describe the results from this large population, and quantitatively compare these with the distributions for other transients with massive star progenitors (LGRBs, Type Ib/c SNe, Type II SNe) using the Kolmogorov–Smirnov (KS) \citep{KS_test} and Anderson–Darling (AD) \citep{AD_test} tests. Both tests are designed to determine whether two distributions arise from the same underlying population, with the AD test being a modified version of the KS test that is more sensitive to the tails of a distribution (whereas the KS test gives more weight to the region around the median of a distribution). While the KS test is more widely used in previous works, we regard the AD test to be a more robust statistical measure, especially in the context of the fractional flux distribution where tail contributions are more prominent.  We provide comparisons of the distributions of physical and host-normalized offsets, galaxy sizes, and fractional flux values, and summarize the KS and AD test $p$-values in Table~\ref{tab:test_pvals}.  We stress that these statistical comparisons were severely limited by the small sample size in the previous study (\citetalias{Lunnan_2015}).

\subsection{Physical Offsets}
\label{sec:physical_offset_distribution}

In Figure~\ref{fig:offsets_mc} we show the cumulative distribution of projected physical offsets ($R_{\rm phys}$). The distribution spans $\approx 0.07 - 6.7$ kpc, with a median of $\langle R_{\rm phys}\rangle\approx 0.73$ kpc. The distribution is overall smooth across the full range of offsets, with no notable gaps. 
The KS and AD tests comparing our distribution with the smaller sample of 16 SLSNe from \citetalias{Lunnan_2015} yield $p$-values of $0.71$ and $0.81$, respectively. This is not surprising given that the sources in the \citetalias{Lunnan_2015} sample are also included in our larger data set. 

We also compare the projected physical offsets to the distributions for LGRBs from \citetalias{Blanchard_2016}, as well as Type Ib/c SNe and Type II SNe from \cite{Kelly_Kirshner_2012}; see Figure~\ref{fig:offsets_mc}. All three populations have systematically larger offsets than SLSNe, LGRBs by a factor 1.4, SNe Ib/c by a factor of 4.4, and SNe II by a factor of 5.5. The KS and AD tests relative to the LGRBs sample yield $p$-values of $0.12$ and $0.20$, respectively, suggesting that the physical offset distributions for SLSNe and LGRBs are consistent with being drawn from the same underlying distribution. For the SNe Ib/c sample, the KS and AD tests yield $p$-values of $2.8\times10^{-11}$ and $2.4\times10^{-10}$, respectively, while for the SNe II the $p$-valuess are $1.3\times10^{-11}$ and $3.6\times10^{-17}$, respectively, thus indicating clearly that the SLSN physical offsets are vastly different from those of SNe Ib/c and II.

The offsets have associated uncertainties, $\sigma_{R_{\rm phys}}$ and $\sigma_{R_{\rm norm}}$, that are dependent on $\sigma_{\rm tie}$, $\sigma_{\rm SN}$, and $\sigma_{\rm host}$. Since the offset is a positive-definite quantity, we cannot assume a Gaussian distribution for its uncertainty. Instead, we use the Rice distribution to represent the probability distribution function \citep{Bloom_2002}:
\begin{equation}
\label{eq:offset_dist}
p(x|R,\sigma_R)=\frac{x}{\sigma_R^2}\exp\left[-\frac{\left(x+R\right)^2}{2\sigma_R^2}\right]I_0\left(\frac{xR}{\sigma_R^2}\right),
\end{equation}
where $R$ and $\sigma_R$ are the offset quantity (physical or normalized) and its uncertainty, and $I_0$ is the zeroth order modified Bessel function of the first kind. Here we employ a Monte Carlo approach with 10,000 iterations to assess the uncertainties on the measured offset distributions. 

Accounting for the uncertainties in the physical offsets, we show in Figure~\ref{fig:offsets_mc} the results of the Monte Carlo simulation by plotting a 2D histogram of the density of points from the cumulative distributions generated at each of the 10,000 iterations. Dark regions indicate a higher density of points, or, in other words, more synthetic distributions from the simulation pass through that region. The median of the distribution of medians from the Monte Carlo simulation is 0.81 kpc with a 90\% confidence interval of $0.71-0.97$ kpc. The overall apparent shift in the Monte Carlo distribution to higher offsets compared to the median distribution is due to the fact that the offset is a positive-definite quantity. Re-calculating the KS and AD tests for each iteration in comparison to LGRBs, SNe Ib/c and SNe II, we confirm the same result as above.

\begin{figure}[t!]
    \centering
    \includegraphics[width=\columnwidth]{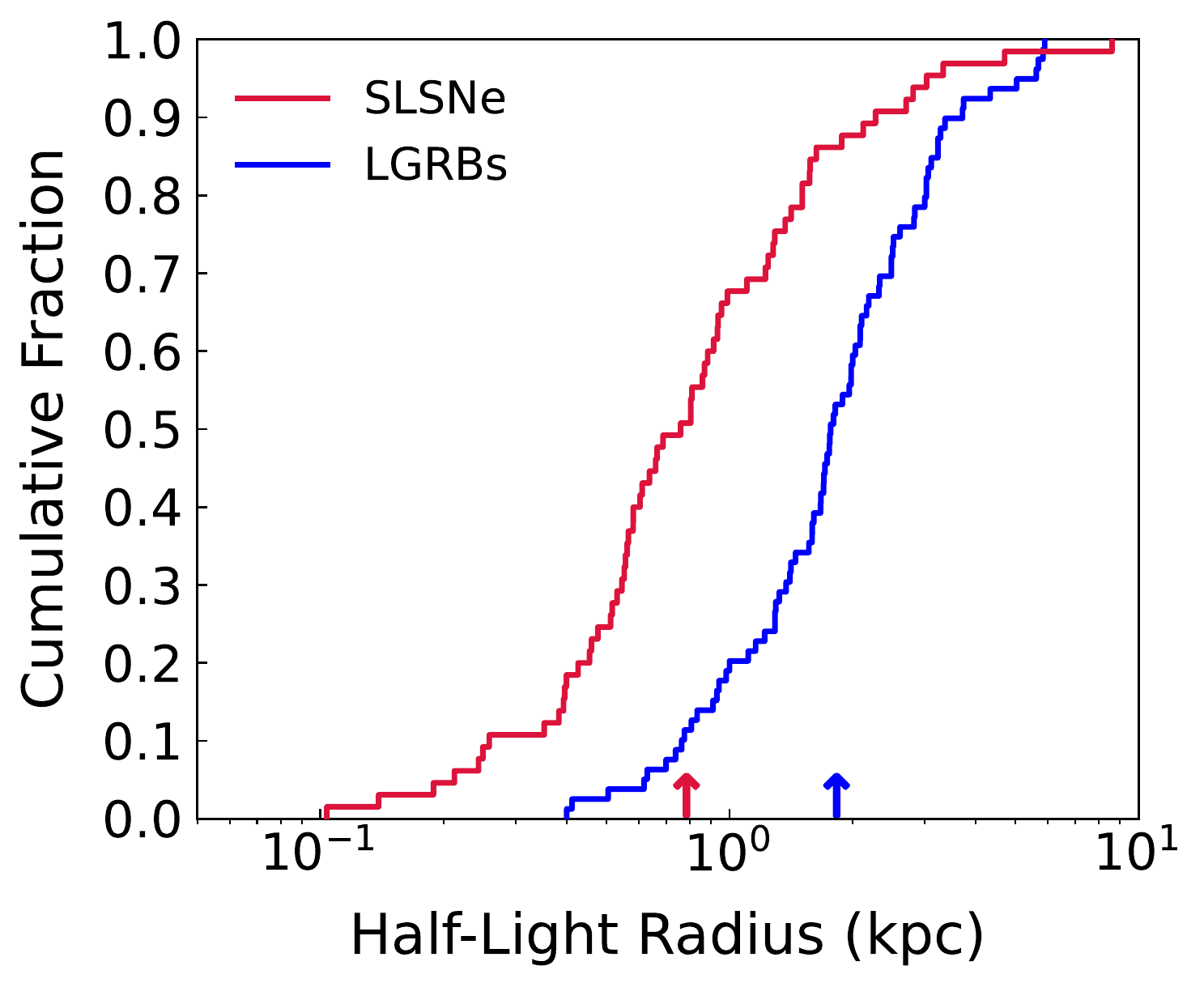}
    \caption{Cumulative distribution of host galaxy half-light radii ($R_{50}$) for our SLSN sample (red), and LGRBs (blue; \citetalias{Blanchard_2016}). The SLSN host galaxies have a median size of $\langle R_{50}\rangle\approx 0.76$ kpc, about a factor of 2.4 times smaller than the LGBR host galaxies. We note that the $R_{50}$ distributions for SNe Ib/c and SNe II were not reported in previous studies.}
    \label{fig:R_h_comp}
\end{figure}

Due to the positive-definite nature of the offsets, offsets with large uncertainties are more likely to be skewed toward higher values. In Appendix~\ref{sec:correction} we explore this issue and undertake a simple procedure to correct for this potential bias.  In Figure~\ref{fig:offsets} we show the ``corrected'' distribution, but find that it overall closely matches the directly observed one.  Since the correction is small, and since it was not applied for other transients to which we compare our SLSN sample, we do not use it in our analysis to prevent introducing additional bias in comparison to previous works.

\begin{figure*}[t!]
    \centering
    \includegraphics[width=0.85\textwidth]{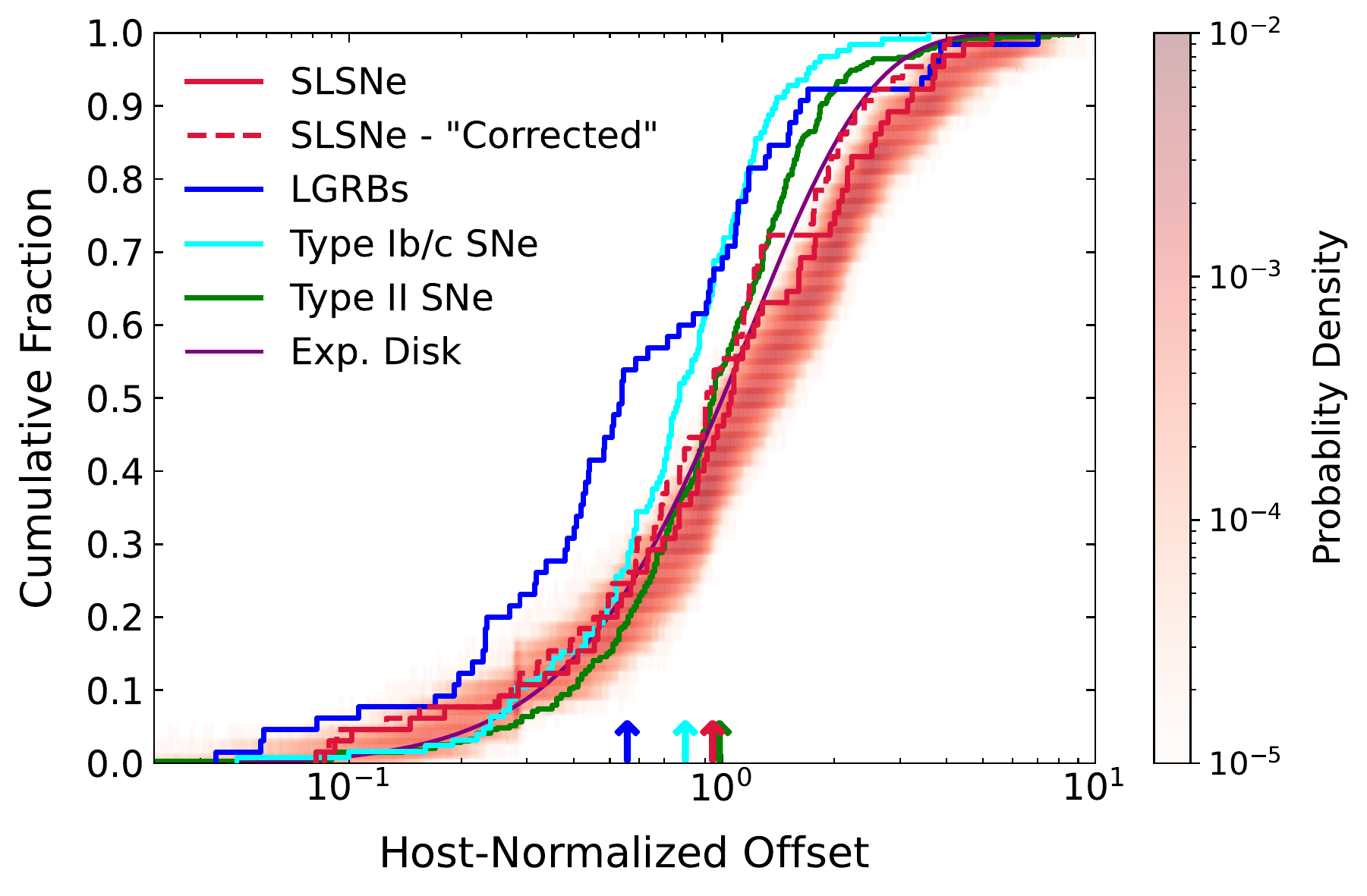}
    \caption{Cumulative distribution of host-normalized offsets for 65 SLSNe from this work (red). Also shown are the distributions for LGRBs (blue;  \citetalias{Blanchard_2016}), SNe Ib/c (cyan) and II (green) SNe from \citet{Kelly_Kirshner_2012}. We also plot the distribution expected for an exponential disk profile (purple). The red shaded region shows the results of our Monte Carlo simulation using the associated uncertainties as a 2D histogram. Arrows at the bottom denote the medians of each distribution. The dashed red line includes a correction factor to account for the positive-definite nature of offsets (see Appendix~\ref{sec:correction} for details).}
    \label{fig:norm_offsets_mc}
\end{figure*}

\begin{figure*}[t!]
    \centering
    \includegraphics[width=0.85\textwidth]{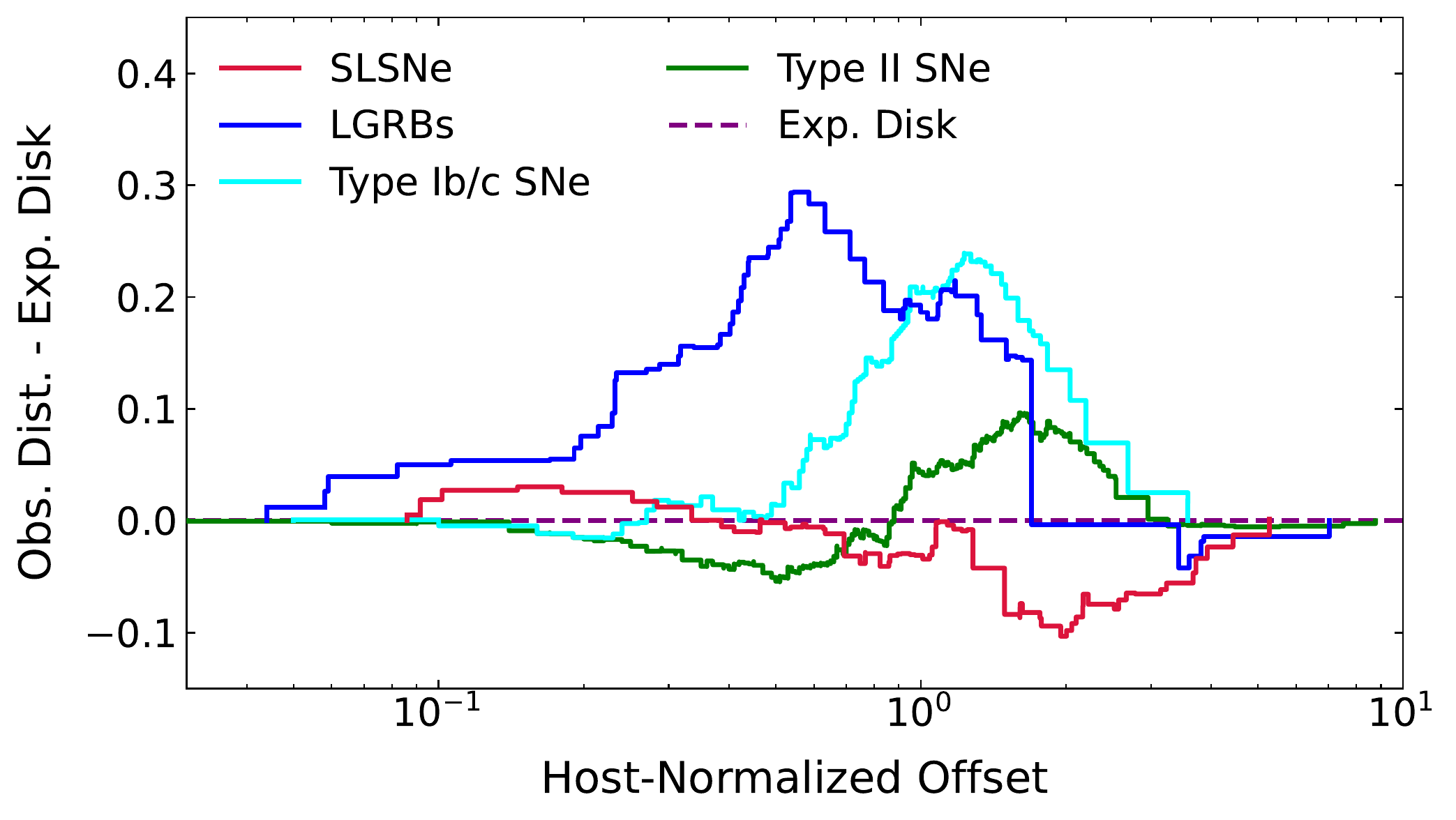}
    \caption{The differences between the cumulative host-normalized offset distributions of our SLSN sample (red), LGRBs (blue), SNe Ib/c (cyan), and SNe II (green) SNe and the cumulative distribution for an exponential disk profile. We find  an overabundance of SLSNe at $R_{\rm norm}\gtrsim 1.5$, compared to an overabundance of LGRBs at $R_{\rm norm}\approx 0.2-1$, and SNe Ib/c and II overabundance at $R_{\rm norm}\approx 1$.}
    \label{fig:exp_disk_ratio}
\end{figure*}

\subsection{Galaxy Sizes}
\label{sec:gal_size}

To compare the projected offsets, both for the SLSN sample itself and in comparison to other transients, in a more meaningful way, we need to normalize each offset by the size of the host galaxy, i.e., $R_{\rm norm}$.  We use $R_{50}$, the circular radius containing half of the galaxy light, to normalize the offsets. We show the distribution of $R_{50}$ for SLSNe in Figure~\ref{fig:R_h_comp}. We find a median value of 0.76 kpc, and an overall range of $\approx 0.1-8.6$ kpc. We also show the $R_{50}$ distribution for LGRBs (\citetalias{Blanchard_2016}), which has a median of 1.8 kpc and a range of $0.4-5.9$ kpc.  Thus, SLSN hosts are on average about a factor of 2.4 times smaller than even the overall compact host galaxies of LGRBs.

The KS and AD tests between the SLSN and LGRB $R_{50}$ distributions yield $p$-values of $4.1\times10^{-9}$ and $3.2\times10^{-11}$, clearly indicating that the two distributions are not drawn from the same underlying population. We only compare the $R_{50}$ distribution to LGRBs, as the values were not reported for SNe Ib/c and SNe II in \citet{Kelly_Kirshner_2012}. 

\begin{figure*}[t!]
    \centering
    \includegraphics[width=0.9\textwidth]{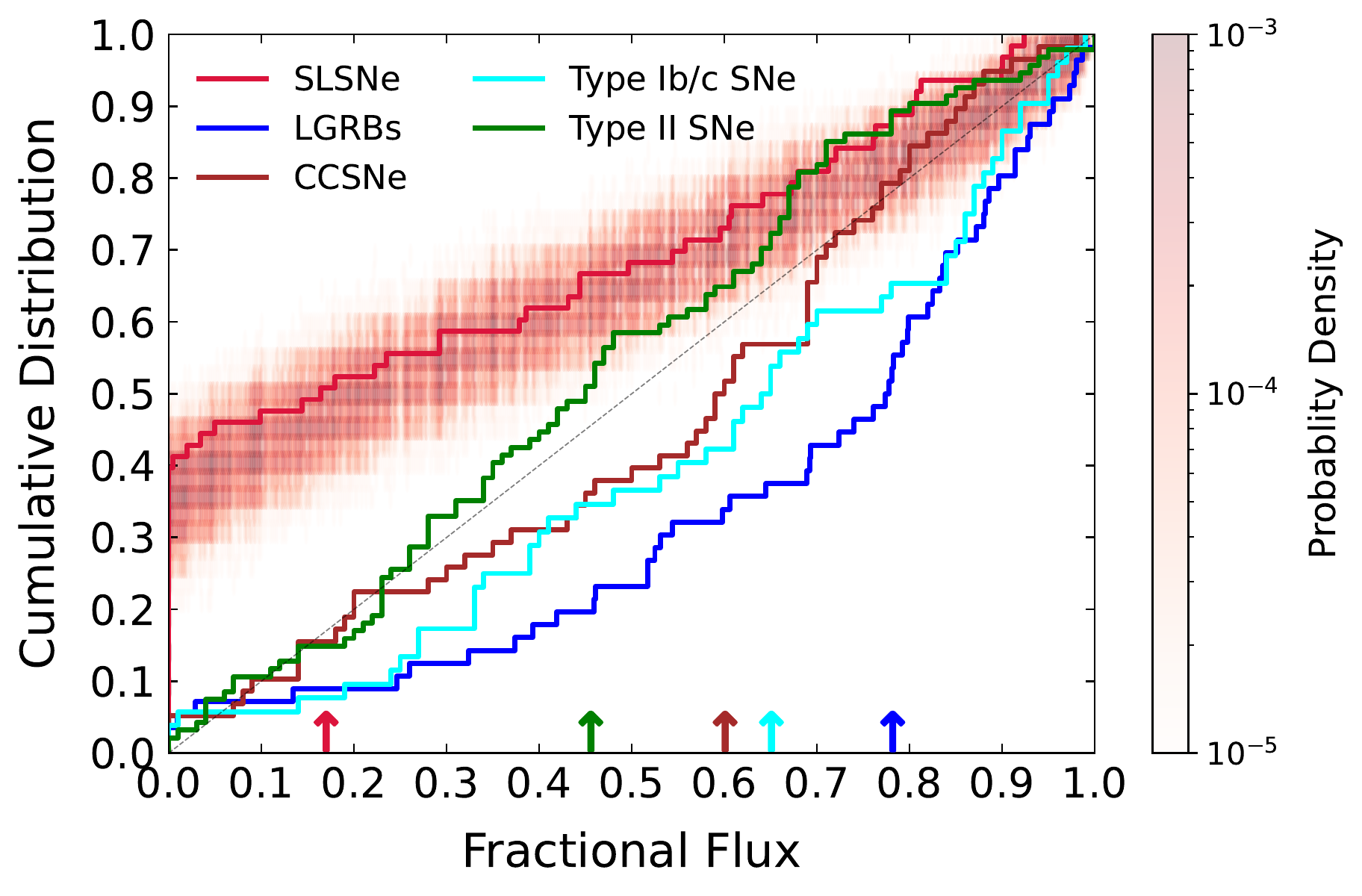}
    \caption{Cumulative distributions of fractional flux for our sample of 63 SLSNe (red; this excludes SN\,2015bn and SN\,2016inl). Also shown are the distributions for LGRBs (blue; \citetalias{Blanchard_2016}), CCSNe (brown; \citealt{Svensson_2010}, SNe Ib/c (cyan) and II (green) from \citet{Kelly_2008}. The red shaded region indicates the $1\sigma$ uncertainty region from our Monte Carlo procedure (\S\ref{sec:frac_flux}). The diagonal dashed line marks the expected fractional flux distribution for a population that uniformly traces the underlying light of their host galaxies. Arrows at the bottom denote the medians of the various distributions.}
    \label{fig:frac_flux_dist}
\end{figure*}

\subsection{Host-Normalized Offsets}
\label{sec:Rnorm_distribution}

In Figure~\ref{fig:norm_offsets_mc} we plot the cumulative distribution of host-normalized offsets ($R_{\rm norm}$). The distribution has a median value of $\approx 1.06$, and spans a range of $\approx 0.09-5.28$ (with about 3/4 of the SLSNe having $R_{\rm norm}\lesssim 2$).  The shaded region in Figure~\ref{fig:norm_offsets_mc} shows the results of our Monte Carlo simulation taking into account the uncertainties on the individual measurements, as described in \S\ref{sec:physical_offset_distribution}. The median of the distribution of medians is $1.17$ with a 90\% confidence interval of $1.01-1.35$. 

Overall, the distribution is reminiscent of an exponential disk profile, the expected surface brightness profile of star-forming disk galaxies, although the SLSN distribution is broader, especially at large offsets. The KS and AD tests comparing the SLSN and exponential disk distributions yield $p$-values of $0.34$ and $0.17$, respectively, indicating that the observed SLSN distribution is consistent with a smooth exponential disk distribution.  Using the Monte Carlo range of cumulative distributions, we find that about $54\%$ and $14\%$ yield KS and AD $p$-values of $\gtrsim 0.05$, respectively.

We also compare the SLSN host-normalized offset distribution to the distributions for LGRBs (\citetalias{Blanchard_2016}), and SNe Ib/c and II \citep{Kelly_Kirshner_2012}. The LGRB and SNe Ib/c distributions have smaller medians by about a factor of 2 and 1.4, respectively, while the SNe II distribution has a comparable median value.  The KS tests comparing the SLSN distribution to the three populations yield $p$-values of $4.1\times10^{-3}$ (LGRBs), $1.1\times10^{-3}$ (SNe Ib/c), and $2.1\times10^{-2}$ (SNe II), while the AD tests yield $p$-values $1.2\times10^{-3}$ (LGRBs), $2.4\times10^{-4}$ (SNe Ib/c), and $1.6\times10^{-2}$ (SNe II), respectively. This indicates that the SLSN host-normalized offset distribution is distinct from those of LGRBs and CCSNe, specifically extending to larger offsets.

To help visualize the comparison between SLSNe and other transients, and each with the exponential disk, in Figure~\ref{fig:exp_disk_ratio} we plot the difference between each cumulative distribution and the exponential disk distribution. The results illustrate that none of these transient populations strictly follow an exponential disk profile, but that SLSNe differ significantly in the way they deviate from the exponential disk distribution.  Specifically, we find that the main deviation for SLSNe is an overabundance of large normalized offsets, $R_{\rm norm}\approx 1.5-4$, while for LGRBs there is an overabundance of small offsets, $R_{\rm norm}\approx 0.2-1$; for SNe Ib/c and II there is an overabundance at $R_{\rm norm}\approx 1$.

\subsection{Fractional Flux Distribution}
\label{sec:frac_flux}

In Figure~\ref{fig:frac_flux_dist} we show the cumulative distribution of fractional flux for 63 SLSN host galaxies\footnote{We exclude SN\,2015bn and SN\,2016inl for which a measurement of the fractional flux is not possible due to the presence of SN emission in the {\it HST} images.}. Also shown is a diagonal 1:1 line which marks the expectation of a population of sources that linearly tracks the underlying light distribution of their host galaxies.  Remarkably, the SLSN sample exhibits a high fraction of events, $\approx 0.4$, with fractional flux value of 0, and has a low median value of $\approx 0.16$. Thus, the locations of SLSNe appear to be significantly skewed to dimmer than average UV regions of their host galaxies.  We also show the resulting 2D probability density using the Monte Carlo procedure described in \S\ref{sec:frac_flux_measurement}; we still find that all 10,000 draws have a substantial fraction of $0.3-0.45$ of the population with fractional flux values of 0.  We note that our distribution is in tension with the smaller sample in \citetalias{Lunnan_2015}, with KS and AD test $p$-values of $1.3\times10^{-2}$ and $3.8\times10^{-4}$, respectively. 

\begin{deluxetable*}{llcccccc}
\tablewidth{\textwidth} 
\tablecaption{Summary of KS and AD Test $p$-values \label{tab:test_pvals}}
\tablehead{\colhead{Test}&\colhead{Parameter} & \colhead{Exp.~Disk} & \colhead{SLSNe (L15)} & \colhead{LGRBs} & \colhead{CCSNe} & \colhead{SNe Ib/c} & \colhead{SNe II}}
\startdata
\multirow{3}{*}{KS}& Physical Offsets & $\cdots$ & 0.71 & $0.12$
 & $\cdots$ & $2.76\times10^{-11}$ & $1.26\times10^{-11}$\\
& Host-Normalized Offsets & 0.34 & 0.53 & $4.04\times10^{-3\phn}$ &  $\cdots$ & $1.12\times10^{-3\phn}$  & $2.11\times10^{-2\phn}$\\
& Fractional Fluxes & $\cdots$ & $1.31\times10^{-2}$ & $1.83\times10^{-6\phn}$ & $4.61\times10^{-5}$ & $5.57\times10^{-6\phn}$ & $5.00\times10^{-6\phn}$\\\hline
\multirow{3}{*}{AD}& Physical Offsets & $\cdots$ & 0.81
& 0.20 & $\cdots$ & $2.39\times10^{-10}$ & $3.60\times10^{-17}$\\
& Host-Normalized Offsets & 0.17 & 0.45 & $1.17\times10^{-3\phn}$ & $\cdots$ & $2.42\times10^{-4\phn}$ & $1.59\times10^{-2\phn}$\\
& Fractional Fluxes & $\cdots$ & $3.78\times10^{-4}$ & $1.39\times10^{-10}$ & $2.06\times10^{-6}$ & $5.23\times10^{-8\phn}$ & $1.25\times10^{-7\phn}$ \\[+2pt]
\enddata
\tablecomments{The comparison samples used here are SLSNe (\citetalias{Lunnan_2015}), LGRBs (\citetalias{Blanchard_2016}), CCSNe \citep{Svensson_2010}, SNe Ib/c and II offsets \citep{Kelly_Kirshner_2012}, and SNe Ib/c and II fractional fluxes \citep{Kelly_2008}.}
\end{deluxetable*}

The results for SLSNe are also remarkable in comparison to the distributions for LGRBs \citepalias{Blanchard_2016}, CCSNe \citep{Svensson_2010}, and SNe Ib/c and II as separate populations \citep{Kelly_2008}.  All of these other populations either roughly track the 1:1 line, or are skewed to {\it brighter} than average regions of their hosts, with median values of about 0.8 (LGRBs), 0.65 (SNe Ib/c), 0.6 (CCSNe), and 0.45 (SNe II); none of the comparison populations exhibit an overabundance of events with fractional flux values of 0.  Formally, the KS tests between the SLSN distribution and the other distributions yield $p$-values of $1.8\times10^{-6}$ (LGRBs), $4.6\times10^{-5}$ (CCSNe), $5.6\times10^{-6}$ (SNe Ib/c), and $5\times 10^{-6}$ (SNe II), respectively.  The AD tests yield $p$-values of $1.4\times10^{-10}$ (LGRBs), $2.1\times10^{-6}$ (CCSNe), $5.2\times10^{-8}$ (SNe Ib/c), and $1.2\times10^{-7}$ (SNe II). Thus, we therefore rule out the possibility that SLSNe are drawn from the same population as other transients in terms of their association with the UV light of their host galaxies.

\begin{figure}[t!]
    \centering
    \includegraphics[width=\columnwidth]{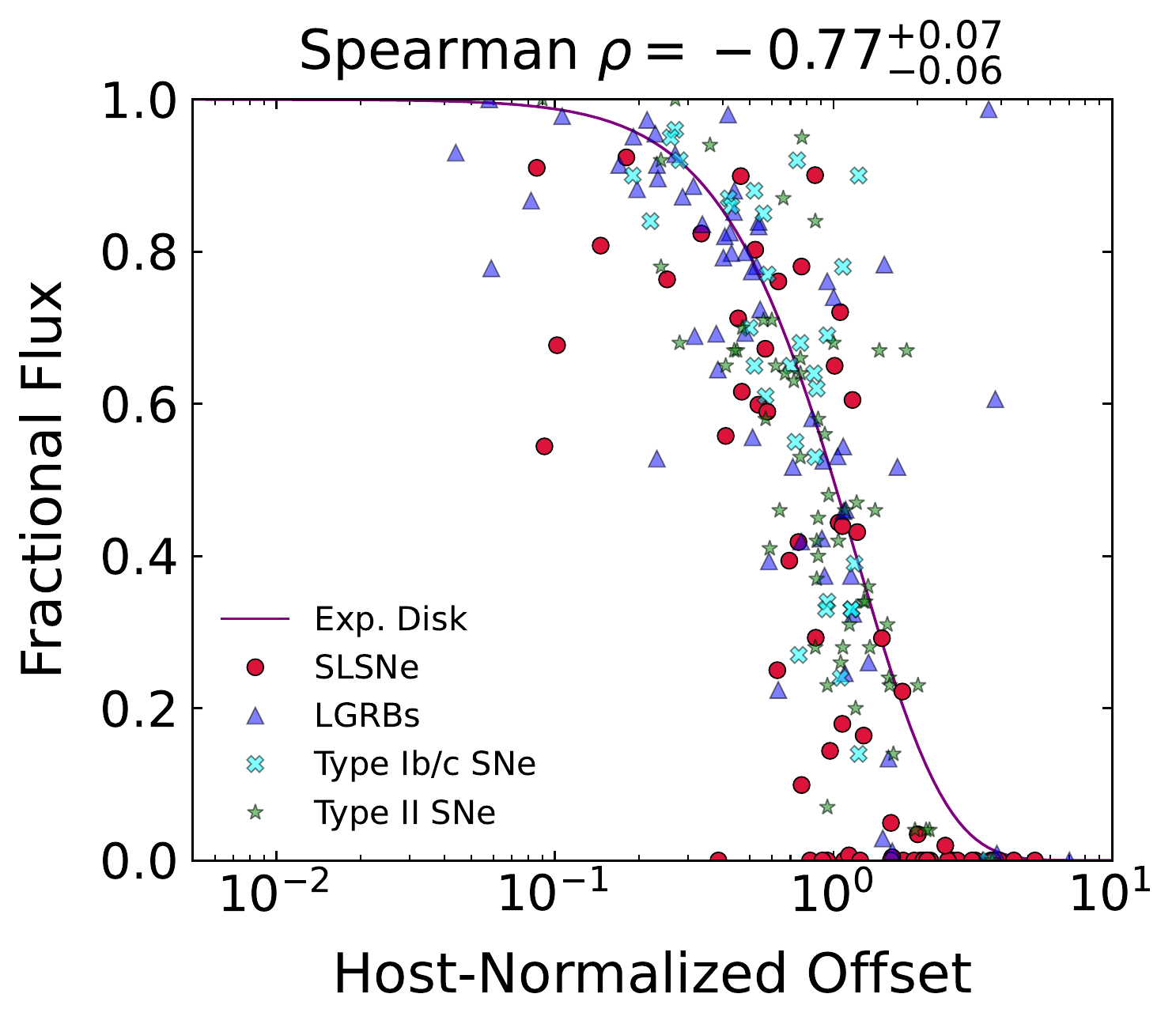}
    \caption{Scatter plots of fractional flux versus host-normalized offsets for SLSNe (red), LGRBs (blue), Type Ib/c SNe (cyan), Type II SNe (green), and predicted relationship for a transient population that follows the exponential disk model exactly (purple). To avoid clutter, we omit showing the associated uncertainties on fractional flux and host-normalized offsets. We find a strong correlation between fractional flux and host-normalized offset for SLSNe, with a median $\rho\approx-0.77$ and low uncertainties.}
    \label{fig:offset_v_fraclight}
\end{figure}

\begin{figure*}[t!]
    \centering
    \includegraphics[width=\textwidth]{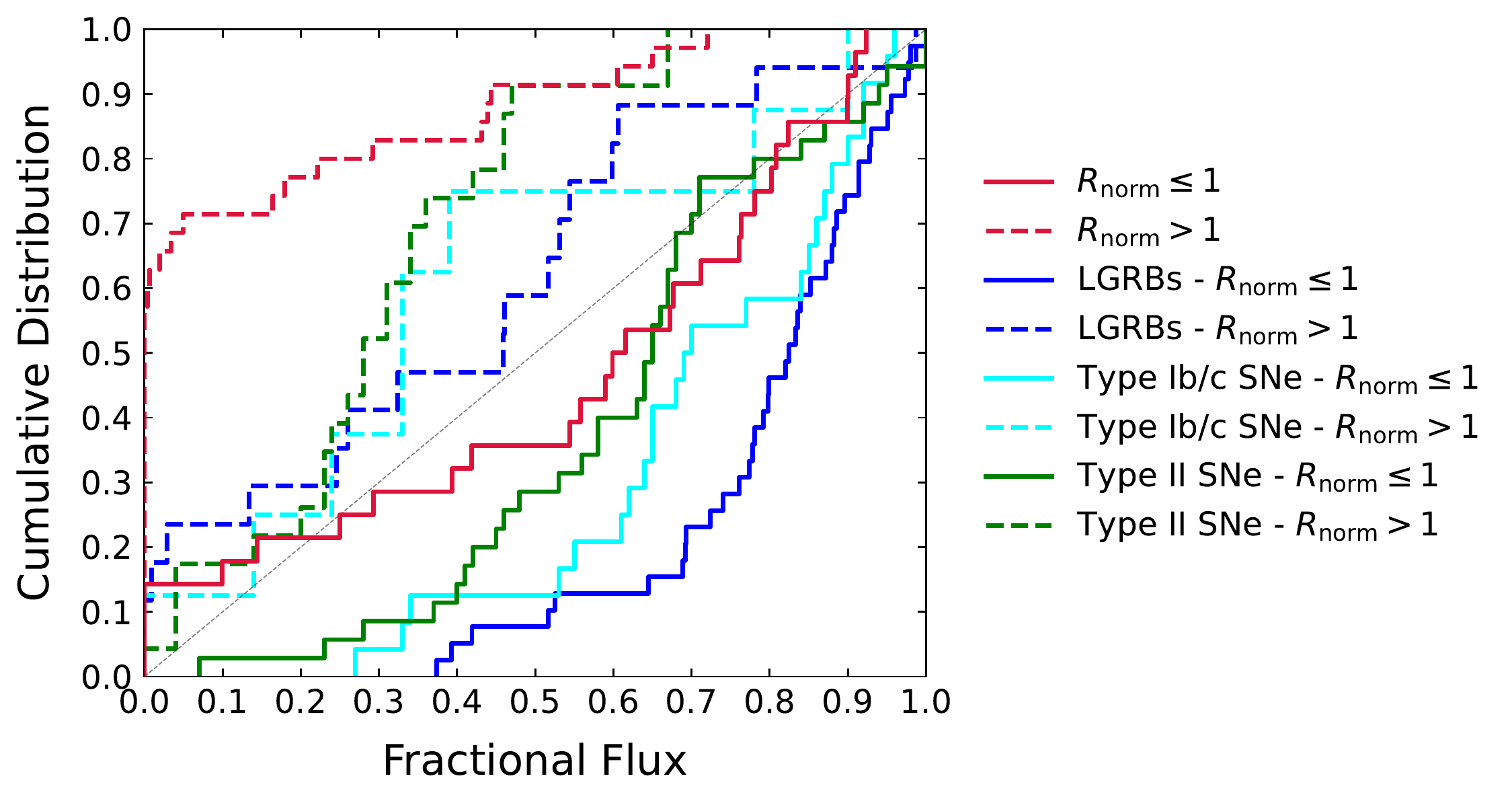}
    \caption{Same as Figure~\ref{fig:frac_flux_dist}, but with the distributions separated into sources with $R_{\rm norm}<1$ (solid) and $R_{\rm norm}\geq 1$ (dashed).  The SLSN population exhibits the lowest fractional flux values in both subsets of the population, but sources with $R_{\rm norm}<1$ roughly track the underlying UV light.}
    \label{fig:frac_flux_norm_cut}
\end{figure*}

\subsection{Fractional Flux-Offset Relationship}

In Figure~\ref{fig:offset_v_fraclight} we plot fractional flux values versus host-normalized offsets for the SLSNe in our sample, LGRBs (\citetalias{Blanchard_2016}), and SNe Ib/c and II (fractional fluxes: \citealt{Kelly_2008}; normalized offsets: \citealt{Kelly_Kirshner_2012}). We also show the relation between fractional flux and normalized offset for an exponential disk distribution, determined by integrating the exponential disk profile at each normalized offset value. We calculate the Spearman's rank correlation coefficients ($\rho$; \citealt{Spearman_1904}) to quantify the strength of correlation between fractional flux and normalized offset, as well as its associated $1\sigma$ bound uncertainty using the method described in \citet{Curran_2014}. We find a clear negative correlation between fractional flux and normalized offsets for SLSNe, with $\rho\approx -0.77$, such that sources with smaller offsets have high fractional flux values. We find similar negative correlations for the other populations ($\rho\approx -0.75$ for LGRBs, $\rho\approx-0.70$ for Type Ib/c SNe, and $\rho\approx-0.76$ for Type II SNe).  Thus, all populations follow a similar trend, which overlaps well with the exponential disk profile. This trend is not surprising given that the central regions of galaxies are brighter. 

However, what does stand out in the SLSN population compared to the other transients (and to an exponential disk) is the unusually large fraction of sources with fractional flux values of 0 and normalized offsets of $\gtrsim 1$.  These can clearly be seen along the bottom of Figure~\ref{fig:offset_v_fraclight}. To further illustrate this point, in Figure~\ref{fig:frac_flux_norm_cut} we plot the cumulative fractional flux distributions for SLSNe and the other transients, separated into sub-populations with $R_{\rm norm}\leq 1$ and $R_{\rm norm}>1$.  We find that SLSNe with $R_{\rm norm}\leq 1$ overall follow the 1:1 line expected for sources the linearly track the UV light distribution of their hosts. However, even within this sub-population the SLSNe are somewhat more skewed to lower fractional values compared to LGRBs and SNe Ib/c and II.  On the other hand, of the SLSNe with $R_{\rm norm}>1$ about 2/3 have fractional flux values of 0, while in the other populations the fraction of such events is $\lesssim 0.15$.

\begin{figure*}[t!]
    \centering
    \includegraphics[width=0.8\textwidth]{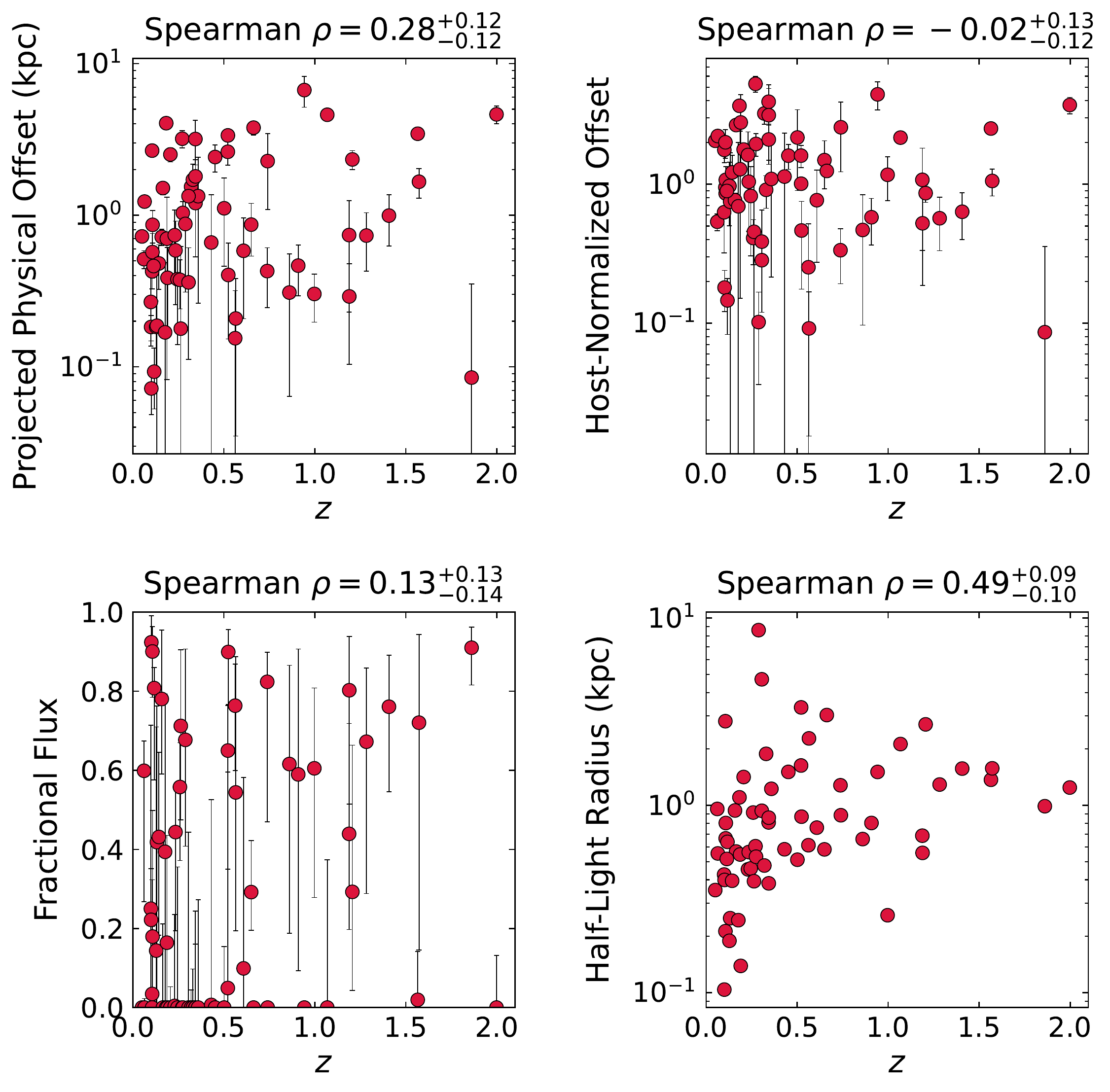}
    \caption{Physical offsets, host-normalized offsets, fractional flux values, and host half-light radii plotted as a function of redshift. In each panel we also list the Spearman rank correlation coefficient.  We find no clear correlations, with the most significant correlation being the one between half-light radius and redshift ($\rho\approx 0.49$).}
    \label{fig:z_v_params}
\end{figure*}

\section{Discussion}
\label{sec:discussion}

The distributions of offsets, fractional fluxes, and host galaxy sizes presented in \S\ref{sec:results} provide the most in-depth analysis of the locations and local environments of SLSNe to date. In this section we explore whether the locations of SLSNe and their host sizes exhibit trends with redshift, as well as any correlations with the inferred properties of their magnetar engines as determined from modeling of the optical light curves. We further explore implications of our results for SLSN progenitors. The Spearman's correlation coefficients are summarized in Table~\ref{tab:spearman}.

\subsection{Trends with Redshift}
\label{sec:z_trends}

In Figure~\ref{fig:z_v_params} we plot the physical and host-normalized offsets, fractional fluxes, and host galaxy half-light radii as functions of SLSN redshift.  No obvious correlation is seen between offset and redshift, with $\rho=0.28^{+0.12}_{-0.12}$ and $\rho=-0.02^{+0.13}_{-0.12}$ for the physical and host-normalized offsets, respectively.  Similarly, no obvious trend is seen between fractional flux and redshift, with $\rho=0.13^{+0.13}_{-0.14}$.  However, we find that nearly all SLSNe with low fractional flux values of $\lesssim 0.2$ are preferentially located at low redshifts, $z\lesssim 0.5$. Finally, we do find a mild correlation between host galaxy half-light radii and redshift, with $\rho=0.49^{+0.09}_{-0.10}$, predominantly due to the prevalence of compact hosts with $R_{50}\lesssim 0.25$ kpc at $z\lesssim 0.25$.

\begin{figure*}[t!]
    \centering
    \includegraphics[width=0.8\textwidth]{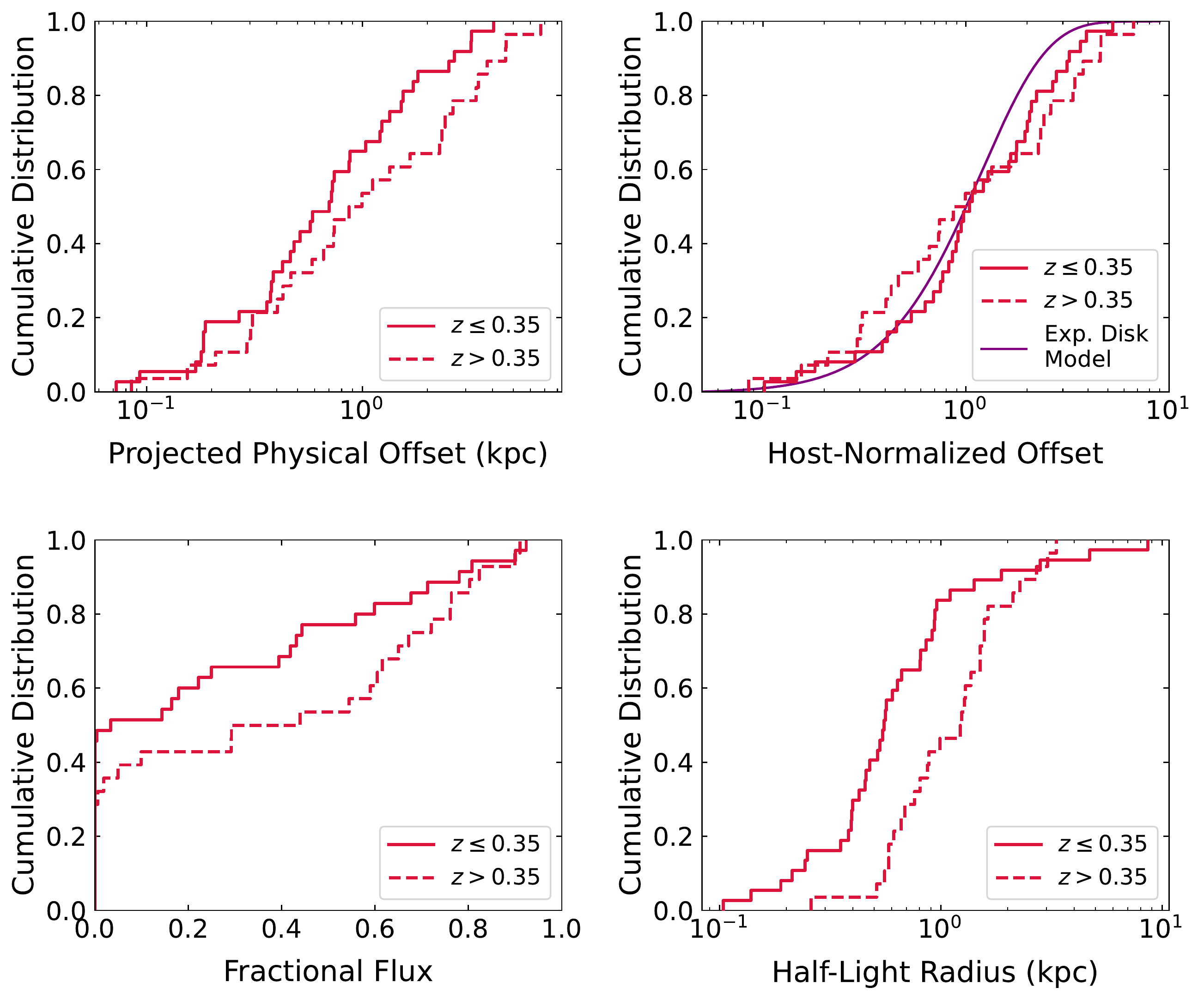}
    \caption{Cumulative distributions of physical offsets, host-normalized offsets, fractional flux values, and host half-light radii binned into two equal-size redshift ranges: $z\le 0.35$ (solid) and $z>0.35$ (dashed). We find a statistically significant trend in half-light radius, and a mild trend in fractional flux.}
    \label{fig:z_binned_dist}
\end{figure*}

To further discern parameter trends with redshift, in Figure~\ref{fig:z_binned_dist} we present cumulative histograms split into two redshift bins at $z=0.35$ (leading to an essentially equal number of $37$ and $38$ SLSNe per bin). We find no clear difference in the cumulative distributions of physical offsets (KS and AD test $p$-values of 0.34 and 0.18) and host-normalized offsets (KS and AD test $p$-values of 0.53 and 0.79). We do find lower fractional flux values at $z\leq 0.35$ (due to the prevalence of sources with fractional flux values of 0 at lower redshifts as noted above); however, the KS and AD test yield $p$-values of 0.27 and 0.12, respectively, indicating no clear statistical difference.  Finally, the only distribution that does exhibit a statistically significant trend is the half-light radius, with KS and AD test $p$-values of $1.4\times10^{-3}$ and $3.5\times10^{-4}$, respectively, indicating that lower redshift SLSN hosts are systematically more compact than at higher redshifts.

\subsection{Trends with Magnetar Engine Parameters}

\begin{figure*}[t!]
    \centering
    \includegraphics[width=\textwidth]{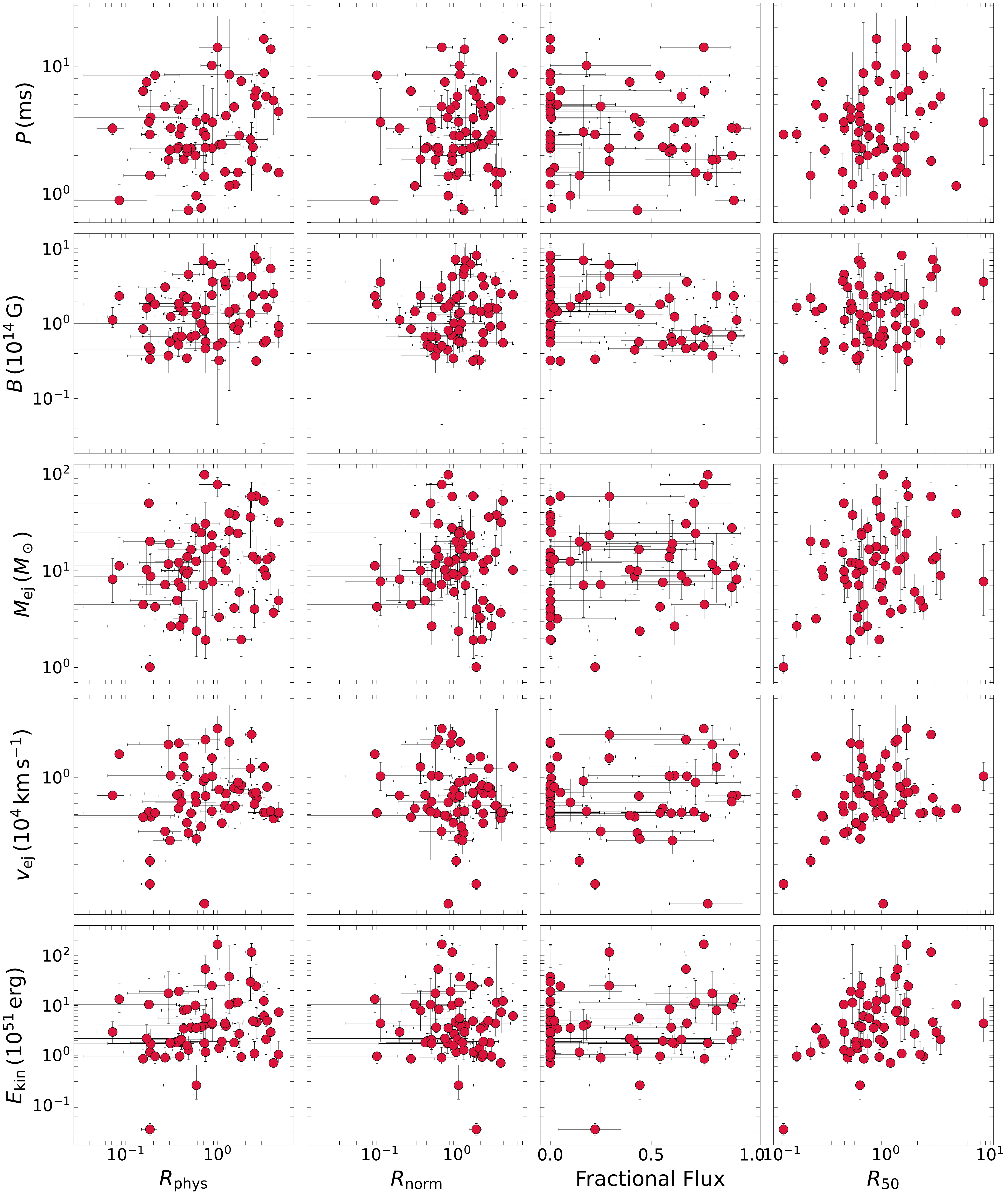}
    \caption{Magnetar model parameters ($P$, $B$, $M_{\rm ej}$, $\varv_{\rm ej}$, and $E_{\rm kin}$), as well as observed peak luminosity ($L_{\rm peak}$) and observed total radiated energy ($E_{\rm rad}$), as a function of physical and host-normalized offsets, fractional flux values, and host half-light radii. The magnetar model parameters are from a uniform study using {\tt MOSFiT} \citep{Gomez_2022}. We do not find any significant correlation between the SLSN parameters and the SLSNe environments.}
    \label{fig:MOSFIT}
\end{figure*}

\begin{figure*}[t!]
    \figurenum{11}
    \centering
    \includegraphics[width=\textwidth]{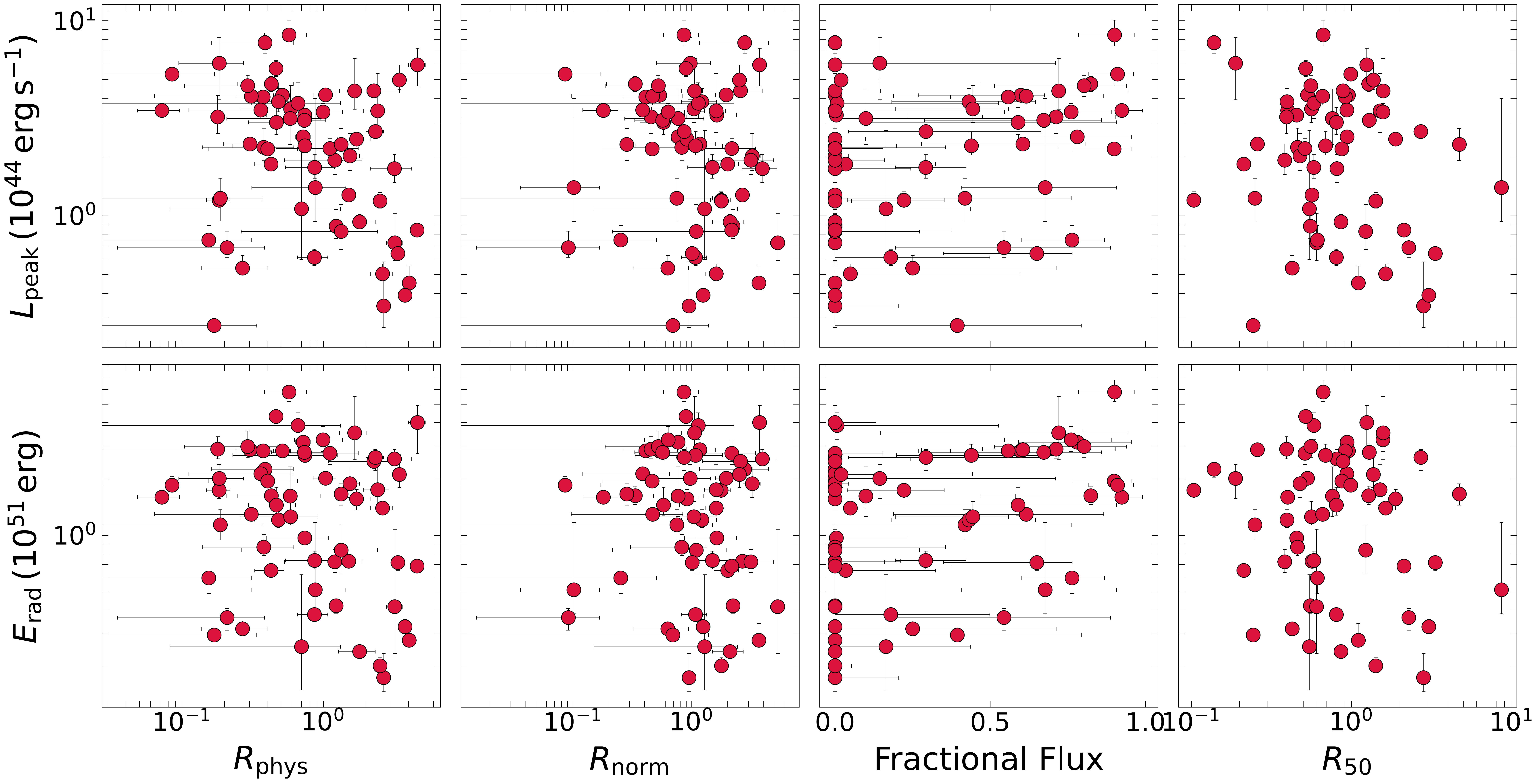}
    \caption{({\it continued})}
\end{figure*}

The optical light curves of SLSNe have been successfully modeled with a magnetar spin-down model \citep{Kasen_Bildsten_2010,Woosley_2010,Dessart_2012,Metzger_2015}, in particular using {\tt MOSFiT}\footnote{https://mosfit.readthedocs.io/en/latest/} \citep{Guillochon_2018}, an open-source, \texttt{Python}-based package that employs a Markov chain Monte Carlo algorithm to fit a one-zone, grey-opacity analytical model to multi-band light curves \citep{Nicholl_2017b,Guillochon_2018}. Here, we compare offset, galaxy size, and fractional flux to magnetar engine parameters to discern possible correlations between SLSN environments and power source. To ensure proper comparison with a consistent model implementation, we use the parameters from \cite{Gomez_2022}.

The four magnetar model parameters we use are the neutron star's initial spin period ($P$) and magnetic field strength ($B$), and the ejecta mass ($M_{\rm ej}$), and velocity ($\varv_{\rm ej}$); we use the latter two to also determine the kinetic energy $E_{\rm kin}=(3/10)M_{\rm ej}\varv_{\rm ej}^2$ of the ejecta. In addition, we also compare to observed peak luminosity ($L_{\rm peak}$) and observed total radiated energy ($E_{\rm rad}$). In Figure~\ref{fig:MOSFIT} we plot the distributions of these 7 SLSN parameters as a function of physical and normalized offsets, fractional flux, and half-light radius.  Visual inspection does not reveal any significant correlation between any pair of parameters. The strongest correlation quantified by Spearman's rank correlation coefficient is between half-light radius and kinetic energy, with a median $\rho\approx 0.31$. Thus, we find no evidence of correlation between SLSN locations and their magnetar engine or explosion properties.

\subsection{Progenitor Implications}

Our key finding is that SLSNe are on average located further away from their galactic centers than LGRBs and CCSNe, and unlike LGRBs and CCSNe, a substantial fraction of $\approx 40\%$ are not correlated with the underlying UV emission of their hosts. The difference relative to LGRBs is particularly intriguing given that both populations represent a rare mode of massive star death ($\lesssim 1\%$ of the overall CCSN rate), and both are thought to be powered by central engines that require rapid rotation (black holes in LGRBs, and magnetars in SLSNe).  Our results, however, indicate that they arise from massive stars that reside in different environments.

Since a strong correlation with bright UV regions, as is the case for LGRBs, can be interpreted as evidence for a particularly young and massive progenitor population (e.g., \citealt{Fruchter_2006, Kelly_2008,Anderson_2012}), the results for SLSNe may be evidence for less massive and somewhat older progenitors.  However, this appears to be in conflict with the inferred pre-explosion mass distribution of SLSNe, which points to systematically more massive progenitors compared to SNe Ib/c (and the small number of LGRBs with inferred progenitor masses; \citealt{Blanchard_2020}).  Similarly, \citet{van_den_Heuvel_2013} proposed a dynamical formation model in young dense star clusters for SLSNe and LGRBs, in which SLSNe are the product of multiple runaway collisions, but our results cast doubt on such a common formation path.

A possible explanation for the substantial fraction of SLSNe occurring outside of UV-bright regions is that their progenitors arise in disrupted binary systems, thereby gaining a velocity kick.  As shown in Figures~\ref{fig:exp_disk_ratio},  \ref{fig:offset_v_fraclight}, and \ref{fig:frac_flux_norm_cut} the dominant contribution to events at low fractional flux is from SLSNe at offsets of $R_{\rm norm}\sim 1-4$, or equivalently $R_{\rm phys}\sim 0.75-3$ kpc. Traveling such distances in the progenitor lifetime span of $\sim 10$ Myr requires velocities of $\gtrsim 10^2$ km s$^{-1}$.  Such high velocities may not be unexpected in models of runaway massive stars from disrupted binary systems, especially for a rare population in predominantly low-metallicity dwarf galaxies (e.g., \citealt{Eldridge_2011}).

The SLSN results are also reminiscent of the isolated locations of luminous blue variables (LBVs) in the Milky Way and Magellanic Clouds \citep{Smith_2015}, explained as possible evidence for LBV formation as mass gainers in binary systems, which are subsequently disrupted. Such a scenario would explain both the large fraction of SLSNe outside of UV-bright regions, and the higher masses of their progenitors at the time of explosion compared to SNe Ib/c. Of course, stars exploding in the LBV phase, with an intact massive hydrogen envelope, cannot be the direct progenitors of SLSNe, but an analogous process involving disruption of a binary system after the progenitor's hydrogen envelope was stripped may be relevant.

Regardless of the exact formation pathway of SLSN progenitors, the offset and fractional flux distributions indicate that factors other than just progenitor mass and metallicity have a significant impact on the formation of SLSNe. However, the details of this pathway do not seem to influence the eventual SN explosion itself, as we do not find any obvious correlation between the SLSN locations and their explosion properties.  This suggests that the explosion properties are mainly governed by the state of the progenitor pre-explosion (e.g., mass, angular momentum).

\begin{deluxetable*}{lcccc}
\tablewidth{\textwidth} 
\tablecaption{Summary of Spearman Rank Correlation Coefficients\label{tab:spearman}}
\tablehead{\colhead{Paramer} & \colhead{$R_{\rm phys}$} & \colhead{$R_{\rm norm}$}  & \colhead{Fractional Flux} & \colhead{$R_{50}$}}

\startdata
Redshift         & $+0.28^{+0.12}_{-0.12}$ & $-0.02^{+0.13}_{-0.12}$ & $+0.13^{+0.13}_{-0.14}$ & $+0.49^{+0.09}_{-0.10}$ \\
Fractional Flux  & \nodata                 & $-0.77^{+0.07}_{-0.05}$ & \nodata                 & \nodata \\
$P$              & $+0.16^{+0.13}_{-0.14}$ & $+0.14^{+0.13}_{-0.13}$ & $-0.22^{+0.13}_{-0.12}$ & $+0.09^{+0.13}_{-0.13}$ \\
$B$              & $+0.22^{+0.12}_{-0.13}$ & $+0.14^{+0.11}_{-0.12}$ & $-0.27^{+0.11}_{-0.11}$ & $+0.11^{+0.13}_{-0.13}$\\
$M_{\rm ej}$     & $+0.24^{+0.11}_{-0.12}$ & $-0.01^{+0.14}_{-0.14}$ & $+0.11^{+0.12}_{-0.13}$ & $+0.24^{+0.12}_{-0.13}$\\
$\varv_{\rm ej}$ & $+0.16^{+0.13}_{-0.13}$ & $-0.06^{+0.12}_{-0.11}$ & $+0.01^{+0.13}_{-0.13}$ & $+0.26^{+0.12}_{-0.13}$\\
$E_{\rm kin}$    & $+0.25^{+0.13}_{-0.14}$ & $-0.06^{+0.13}_{-0.13}$ & $-0.11^{+0.13}_{-0.13}$ & $+0.31^{+0.11}_{-0.12}$\\
$L_{\rm peak}$   & $-0.25^{+0.15}_{-0.15}$ & $-0.16^{+0.13}_{-0.13}$ & $-0.28^{+0.13}_{-0.13}$ & $-0.13^{+0.14}_{-0.13}$\\
$E_{\rm rad}$    & $-0.15^{+0.14}_{-0.13}$ & $-0.13^{+0.12}_{-0.12}$ & $-0.32^{+0.12}_{-0.13}$ & $-0.04^{+0.13}_{-0.12}$\\[+2pt]
\enddata
\end{deluxetable*}

\section{Conclusions}
\label{sec:conclusion}

We have carried out the most comprehensive study of the locations of SLSNe within their host galaxies to date using archival {\it HST} data for 65 SLSNe. We determine both the offset and fractional flux distributions for the sample, and compare these to other transients with massive star progenitors (LGRBs and CCSNe).  Our key findings are as follows:

\begin{enumerate}
\item SLSN host galaxies are more compact than the host galaxies of LGRBs (median of 0.76 versus 1.8 kpc) at large statistical significance (KS test $p$-value of $4.1\times10^{-9}$).

\item The physical offsets of SLSNe have a median of about 0.73 kpc (90\% confidence interval of $0.71-0.97$ kpc from a Monte Carlo simulation). These offsets are systematically smaller than for LGRBs and CCSNe.

\item The host-normalized offsets of SLSNe have a median of 1.06 (in units of $R_{50}$); 90\% confidence interval of $1.01-1.35$ from a Monte Carlo simulation). The distribution overall traces an exponential disk profile, but with a statistically significant overabundance of sources at $R_{\rm norm}\gtrsim 1.5$.  The host-normalized offsets are systematically larger than for LGRBs and CCSNe, with the closest match being SNe II.

\item The fractional flux distribution of SLSNe has a low median value of 0.16, with about 40\% of the sources being located in regions with fractional flux values of 0.  This distribution is strikingly different from those of LGRBs (with a preference for bright UV regions, median value of about 0.8) and CCSNe.

\item The host-normalized offsets of SLSNe strongly correlate with their fractional fluxes, in a similar manner to those of LGRBs and CCSNe; the main distinction is the overabundance of SLSNe with fractional flux values of 0 and normalized offsets of $R_{\rm norm}\gtrsim 1$.  We find that SLSNe with $R_{\rm norm}\le 1$ have a fractional flux distribution that linearly tracks the underlying UV light of their hosts, but about 60\% of those with $R_{\rm norm}>1$ have fractional flux values of 0 in clear distinction from LGRBs and CCSNe with $R_{\rm norm}>1$.

\item The physical and normalized offsets, fractional flux values, and host half-light radii do not show statistically significant trends with redshift. When binned into low ($z\le 0.35$) and high ($z>0.35$) redshift ranges, however, the half-light radius distributions are statistically different, with lower redshift SLSN hosts being systematically more compact.

\item There is no significant correlation between the locations of SLSNe and their explosion and magnetar engine parameters.

\item The substantial difference in SLSN and LGRB locations indicates that while both are rare classes of CCSNe most likely powered by central engines, their progenitors follow different formation pathways.

\item The large fraction of SLSNe outside of UV-bright regions may point to progenitors formed as runaway stars from disrupted binary systems with kick velocities of $\sim 10^2$ km s$^{-1}$.

\end{enumerate}

With the upcoming Vera C.~Rubin Observatory Legacy Survey of Space and Time, we expect a substantial increase in the SLSN discovery rate, extending to higher redshifts than at the present (e.g., \citealt{Villar_2018}).  Studies of this larger SLSN population with {\it HST} and {\it JWST} will be critical for exploring redshift trends, and perhaps subtle correlation between SLSN environments and their explosion properties that cannot be discerned in the current sample.

\acknowledgments
We thank Josh Grindlay, Daichi Hiramatsu, Ashley Villar, and Irwin Shapiro for helpful discussions and comments. We thank Mark Huber for providing images for PS1-MDS SLSNe, Charlotte Angus for providing images for a subset of DES SLSNe, and Matt Nicholl for supplying {\tt galfit} models for SN\,2015bn. The Berger Time Domain group at Harvard is supported in part by NSF and NASA grants, including support by the NSF under grant AST-2108531, as well as by the NSF under Cooperative Agreement PHY-2019786 (The NSF AI Institute for Artificial Intelligence and Fundamental Interactions http://iafi.org/). P.K.B.~is supported by a CIERA Postdoctoral Fellowship. S.G.~is supported by an STScI Postdoctoral Fellowship. 

This research is based in part on observations made with the NASA/ESA Hubble Space Telescope obtained from the Space Telescope Science Institute, which is operated by the Association of Universities for Research in Astronomy, Inc., under NASA contract NAS 5–26555. These observations are associated with programs GO-9500, GO-12529, GO-12786, GO-13022, GO-13025, GO-13326, GO-13858, GO-14743, GO-15140, GO-15162, GO-15303, GO-15496, GO-16239, GO-16657, and GO-17181. 

The Pan-STARRS1 Surveys have been made possible through contributions of the Institute for Astronomy, the University of Hawaii, the Pan-STARRS Project Office, the Max-Planck Society and its participating institutes, the Max Planck Institute for Astronomy, Heidelberg and the Max Planck Institute for Extraterrestrial Physics, Garching, The Johns Hopkins University, Durham University, the University of Edinburgh, Queen's University Belfast, the Harvard-Smithsonian Center for Astrophysics, the Las Cumbres Observatory Global Telescope Network Incorporated, the National Central University of Taiwan, the Space Telescope Science Institute, the National Aeronautics and Space Administration under Grant No. NNX08AR22G issued through the Planetary Science Division of the NASA Science Mission Directorate, the National Science Foundation under Grant No. AST-1238877, the University of Maryland, and Eotvos Lorand University (ELTE).

This research has made use of the NASA/IPAC Infrared Science Archive, which is funded by the National Aeronautics and Space Administration and operated by the California Institute of Technology. 

This research uses services or data provided by the Astro Data Lab at NSF’s NOIRLab. NOIRLab is operated by the Association of Universities for Research in Astronomy (AURA), Inc. under a cooperative agreement with the National Science Foundation.

This project used public archival data from the Dark Energy Survey (DES) as distributed by the Astro Data Archive at NSF's NOIRLab. Funding for the DES Projects has been provided by the US Department of Energy, the US National Science Foundation, the Ministry of Science and Education of Spain, the Science and Technology Facilities Council of the United Kingdom, the Higher Education Funding Council for England, the National Center for Supercomputing Applications at the University of Illinois at Urbana-Champaign, the Kavli Institute for Cosmological Physics at the University of Chicago, Center for Cosmology and Astro-Particle Physics at the Ohio State University, the Mitchell Institute for Fundamental Physics and Astronomy at Texas A\&M University, Financiadora de Estudos e Projetos, Fundação Carlos Chagas Filho de Amparo à Pesquisa do Estado do Rio de Janeiro, Conselho Nacional de Desenvolvimento Científico e Tecnológico and the Ministério da Ciência, Tecnologia e Inovação, the Deutsche Forschungsgemeinschaft and the Collaborating Institutions in the Dark Energy Survey. The Collaborating Institutions are Argonne National Laboratory, the University of California at Santa Cruz, the University of Cambridge, Centro de Investigaciones Enérgeticas, 22 Medioambientales y Tecnológicas- Madrid, the University of Chicago, University College London, the DES-Brazil Consortium, the University of Edinburgh, the Eidgenössische Technische Hochschule (ETH) Zürich, Fermi National Accelerator Laboratory, the University of Illinois at Urbana-Champaign, the Institut de Ciències de l’Espai (IEEC/CSIC), the Institut de Física d’Altes Energies, Lawrence Berkeley National Laboratory, the Ludwig-Maximilians Universität München and the associated Excellence Cluster Universe, the University of Michigan, the  NSF’s NOIRLab, the University of Nottingham, the Ohio State University, the OzDES Membership Consortium, the University of Pennsylvania, the University of Portsmouth, SLAC National Accelerator Laboratory, Stanford University, the University of Sussex, and Texas A\&M University.

The Liverpool Telescope is operated on the island of La Palma by Liverpool John Moores University in the Spanish Observatorio del Roque de los Muchachos of the Instituto de Astrofisica de Canarias with financial support from the UK Science and Technology Facilities Council.

Based in part on observations obtained at the international Gemini Observatory, a program of NSF’s NOIRLab, which is managed by the Association of Universities for Research in Astronomy (AURA) under a cooperative agreement with the National Science Foundation on behalf of the Gemini Observatory partnership: the National Science Foundation (United States), National Research Council (Canada), Agencia Nacional de Investigaci\'{o}n y Desarrollo (Chile), Ministerio de Ciencia, Tecnolog\'{i}a e Innovaci\'{o}n (Argentina), Minist\'{e}rio da Ci\^{e}ncia, Tecnologia, Inova\c{c}\~{o}es e Comunica\c{c}\~{o}es (Brazil), and Korea Astronomy and Space Science Institute (Republic of Korea).

This paper includes data gathered with the 6.5-m Magellan Telescopes located at Las Campanas Observatory, Chile.  Observations reported here were obtained at the MMT Observatory, a joint facility of the University of Arizona and the Smithsonian Institution. This work has made use of data from the European Space Agency (ESA) mission Gaia (https://www.cosmos.esa.int/gaia), processed by the Gaia Data Processing and Analysis Consortium (DPAC, https://www.cosmos.esa.int/
web/gaia/dpac/consortium). Funding for the DPAC has been provided by national institutions, in particular the institutions participating in the Gaia Multilateral Agreement.

\vspace{12pt} 

\facility{Blanco (DECam), Gemini: South (GMOS), {\it HST} (ACS, WFC3), MMT(Binospec), PS1, IPTF, Magellan: Baade (IMACS), PTF, ZTF}

\vspace{12pt}

\textit{Software}: AstroDrizzle \citep{Gonzaga_2012}, Astropy \citep{Astropy_2013,Astropy_2018}, Matplotlib \citep{Hunter_2007}, NumPy \citep{Oliphant_2006}, Photutils \citep{Bradley_2022}, pymccorrelation \citep{Curran_2014,Privon_2020}, PyZOGY \citep{Zackay_2016}, Scipy \citep{Virtanen_2020}

\clearpage
\appendix 

\section{Host Galaxy Non-detections}
\label{sec:no_host}

\begin{figure*}[t!]
    \centering
    \includegraphics[width=\textwidth]{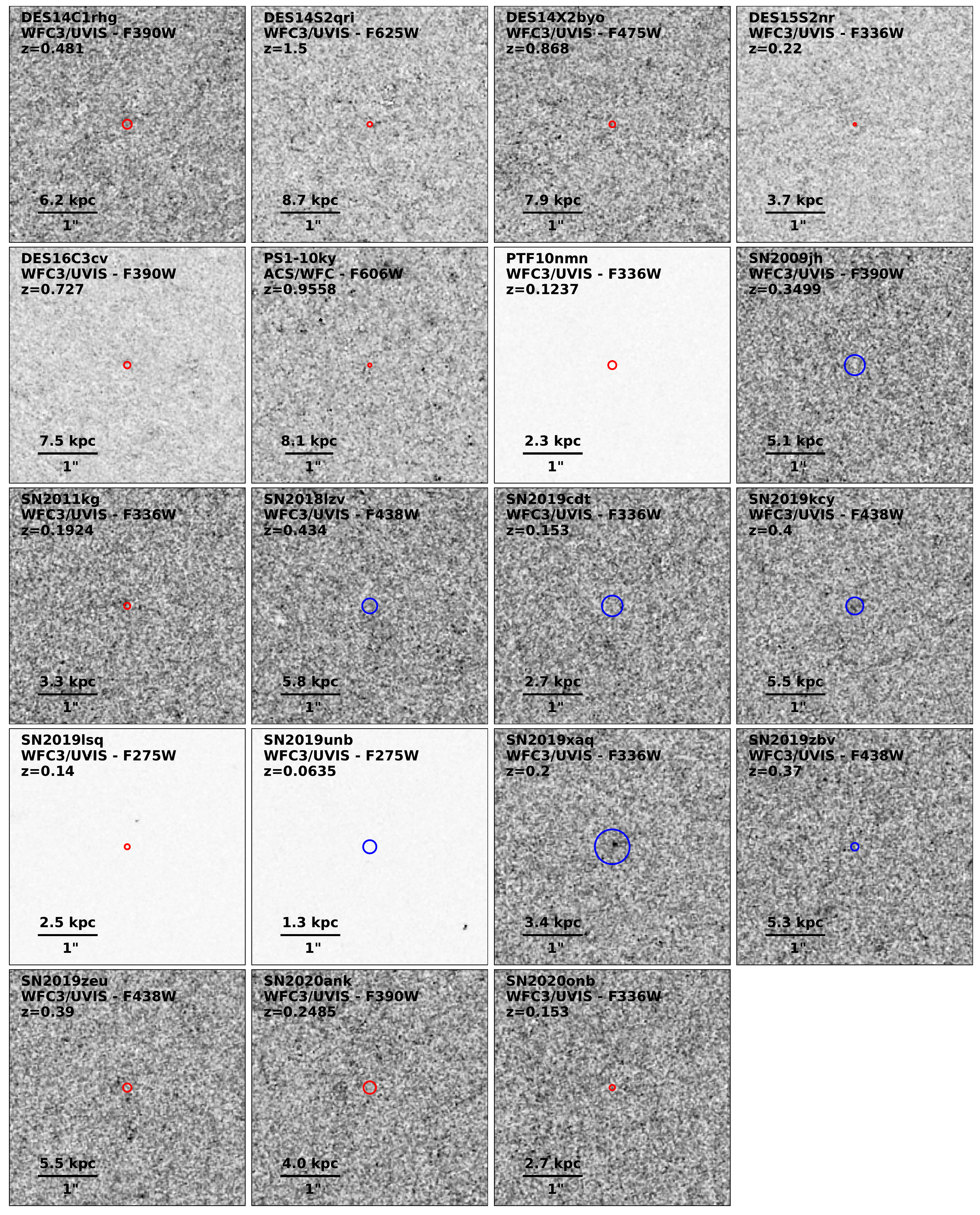}
    \caption{{\it HST} drizzled images of 19 SLSNe with available SN imaging and successful astrometric alignment, but no detected host. The images are centered on the centroid location of the SLSNe and aligned with North up and East to the left. Solid circles indicating the location of the SLSNe, with a radius corresponding to $1\sigma$ uncertainty. Red and blue circles indicate positions determined using relative or absolute astrometry, respectively.}
    \label{fig:offsets_nogal}
\end{figure*}

In Figure~\ref{fig:offsets_nogal} we show the {\it HST} images for the 19 SLSNe without a detected host galaxy.

\section{Offset Corrections}
\label{sec:correction}

\begin{figure*}[t!]
    \centering
    \includegraphics[width=0.8\textwidth]{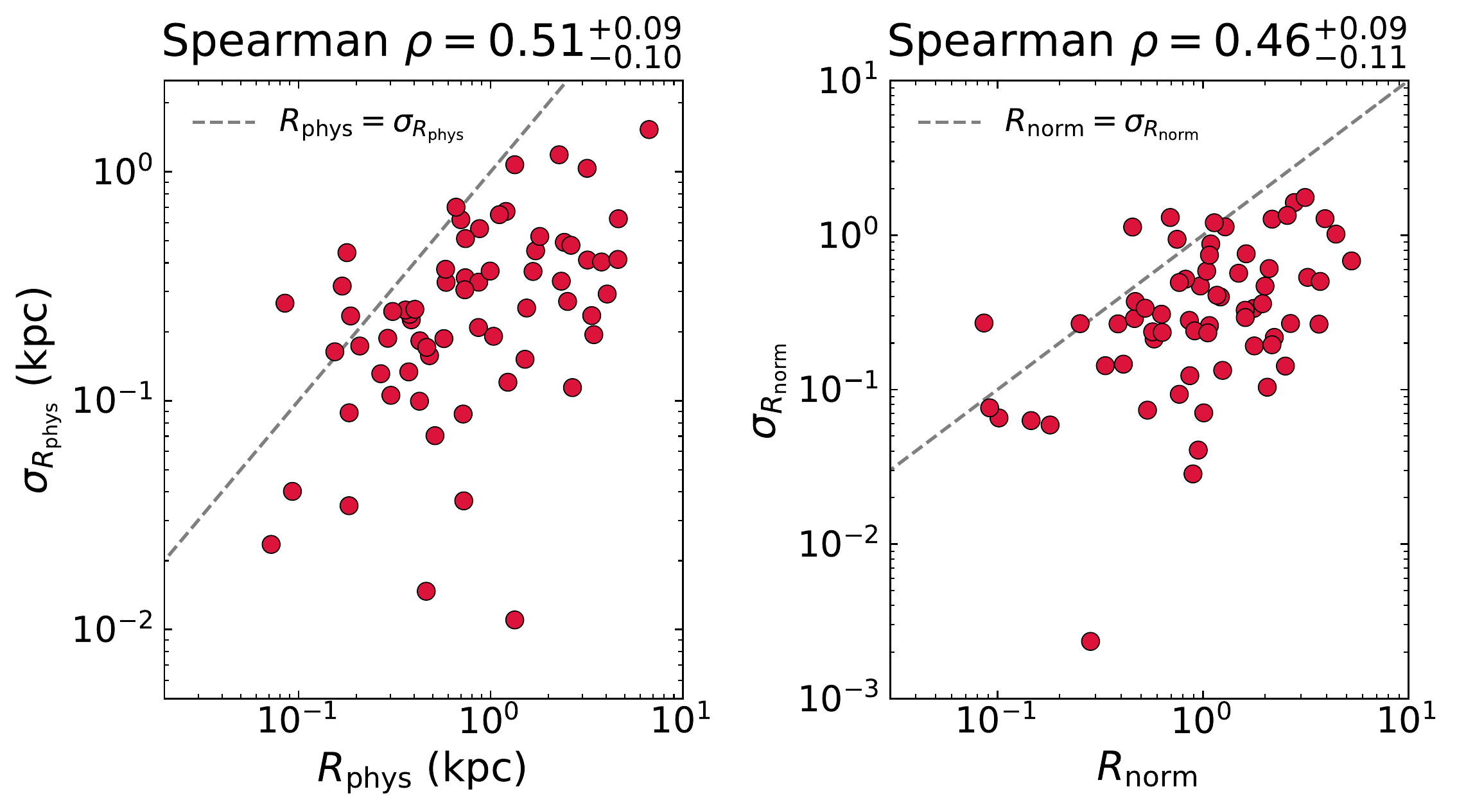}
    \caption{{\it Left}: Physical offset uncertainty ($\sigma_{R_{\rm phys}}$) versus physical offset ($R_{\rm phys}$). {\it Right}: Host-normalized offset uncertainty ($\sigma_{R_{\rm norm}}$) versus host-normalized offset ($R_{\rm norm}$). The dashed line in each panel indicates the 1:1 line where the offset is equal to its uncertainty. The dearth of sources with small offsets and large uncertainties (the upper left quadrant of each panel) is due to the positive-definite nature of the offset quantity.}
    \label{fig:R_v_sigmaR}
\end{figure*}

Since offsets are a positive-definite quantity, sources with a large offset uncertainty may also be systematically skewed to larger offset values.  Since the offset uncertainty $\sigma_R$ ($\sigma_{R_{\rm phys}}$ or $\sigma_{R_{\rm norm}}$) is not a physical property related to the offset $R$ ($R_{\rm phys}$ or $R_{\rm norm}$) itself, but rather is solely due to measurement uncertainties, there should be no correlation between $R$ and $\sigma_R$. In Figure~\ref{fig:R_v_sigmaR} we plot 
$\sigma_{R_{\rm phys}}$ versus $R_{\rm phys}$ and $\sigma_{R_{\rm norm}}$ versus $R_{\rm norm}$, and find a mild correlation, which is specifically due to the apparent lack of sources with large uncertainties and small offsets; this is exactly the systematic effect described above. 

\begin{figure*}[t!]
    \centering
    \includegraphics[width=\textwidth]{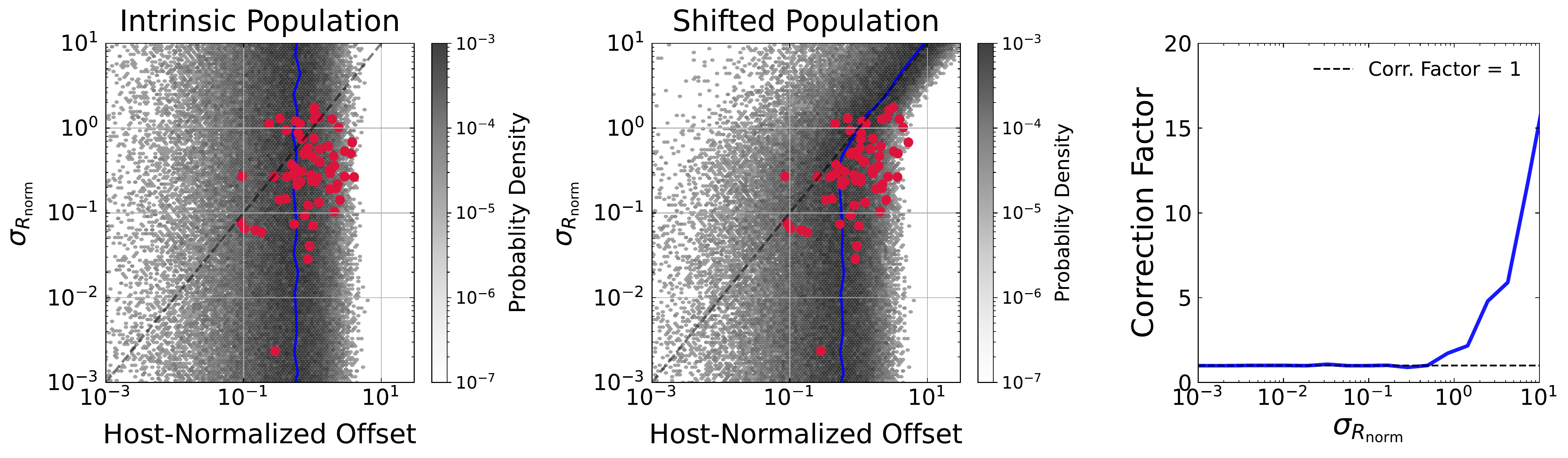}\hfill
    \caption{{\it Left}: 2D probability density plot showing the host-normalized offsets for a population drawn from an exponential disk profile with randomly assigned uncertainties. The blue line marks the peak value as a function of $\sigma_{R_{\rm norm}}$.  Plotted in red are the ``corrected'' $R_{\rm norm}$ values for SLSNe. {\it Middle}: 2D probability density plot showing the host-normalized offsets after sampling with the Rice distribution. The blue line marks the peak value as a function of $\sigma_{R_{\rm norm}}$; the distribution is clearly skewed to larger values of $R_{\rm norm}$ at larger $\sigma_{R_{\rm norm}}$. Plotted in red are the measured values for SLSNe. {\it Right:} The mean correction factor as a function of $\sigma_{R_{\rm norm}}$ is calculated as the ratio of the peaks (blue lines) between the two populations.}
    \label{fig:rice_effect}
\end{figure*}

To quantify the impact of this effect we experiment with a simple procedure to determine a correction factor for each host-normalized offset value, using the exponential disk distribution.  First, we draw 100,000 random offset values from the exponential disk profile and assign to each an  uncertainty drawn from a log-uniform distribution spanning $\sigma_{R_{\rm norm}}=10^{-3}-10$. This provides an input ``intrinsic'' population; see left panel of Figure~\ref{fig:rice_effect}. Second, we reconstruct a Rice distribution for each offset-uncertainty data point using Equation~\ref{eq:offset_dist}, and randomly draw a point from the distribution. This ``shifted'' population (middle panel of Figure~\ref{fig:rice_effect}) is empirically equivalent to what we measure and report in \S\ref{sec:physical_offset_distribution} and \S\ref{sec:Rnorm_distribution}. Lastly, we track the peak of the distribution at each uncertainty value for both the intrinsic and shifted populations, and use the resulting ratio to determine a mean correction factor as a function of $\sigma_{R_{\rm norm}}$; see right panel of Figure~\ref{fig:rice_effect}.  This correction factor is $\gtrsim 1$ only at $\sigma_{R_{\rm norm}}\gtrsim 1$. Finally, we apply this correction factor to the observed SLSN offset distributions, as shown by the dashed lines in Figures~\ref{fig:offsets_mc} and \ref{fig:norm_offsets_mc}. Overall, we find that this correction is rather minimal and does not affect the results of our study; the corrected $R_{\rm norm}$ distribution has a slightly lower median of 0.93 versus 1.06 for the measured distribution. We also note that none of the other comparison samples (LGRBs, SNe) have been subjected to this analysis so it is likely that a correction would shift all populations by a comparable amount.  

\clearpage
\bibliography{Reference}
\bibliographystyle{aasjournal}

\end{document}